\documentclass{ifacconf}
\usepackage{amsmath}
\usepackage{amssymb}
\usepackage{subcaption}

\usepackage{braket}
\usepackage{color}

\newtheorem{definition}{Definition}
\newtheorem{example}{Example}

\usepackage{graphicx}      
\usepackage[dvipsnames]{xcolor}
\usepackage{natbib}        

\newcommand{\uu}{\mathbf{u}}
\newcommand{\x}{\mathbf{x}}

\definecolor{amber}{rgb}{1.0, 0.49, 0.0}

\begin{document}
	\begin{frontmatter}
		
		\title{Compositional Approximately Bisimilar Abstractions of Interconnected Systems\thanksref{footnoteinfo}} 
		
		\thanks[footnoteinfo]{This work was partly supported by the Google Research Grant, the SERB Start-up Research Grant, the CSR Grant by Nokia Corporation and the ANR PIA funding: ANR-20-IDEES-0002.}
		
		\author[First]{Belamfedel Alaoui Sadek} 
		\author[Second]{Saharsh} 
		\author[Second]{Pushpak Jagtap}
		\author[Third]{Adnane Saoud}
		
		\address[First]{LISAC, faculty of sciences dhar el mehraz university sidi mohamed ben abdellah. (e-mail: sadek.belamfedelalaoui@usmba.ac.ma)}
		\address[Second]{Robert Bosch Center for Cyber-Physical Systems, Indian Institute of Science, 
			Bangalore, India (e-mail: \{saharsh2021,pushpak\}@iisc.ac.in)}
		\address[Third]{ CentraleSup\'elec, Universit\'e Paris Saclay, Gif-sur-Yvette, France. (e-mail: adnane.saoud@centralesupelec.fr)}
		
		\begin{abstract}                
			This paper formulates and studies the concepts of approximate (alternating) bisimulation relations characterizing equivalence relations between interconnected systems and their abstractions. These equivalence relations guarantee that the abstraction conserves the original model's dynamics. We develop a compositional approach for abstraction-based controller synthesis by relying on the notions of approximate composition and incremental input-to-state stability. In particular, given a large-scale system consisting of interconnected components, we provide conditions under which the concept of approximate (alternating) simulation relation is preserved when going from the subsystems to the large-scale interconnected system. The effectiveness of the proposed results has been evaluated through traffic congestion control.
		\end{abstract}
		
		\begin{keyword}
			Symbolic control, Compositional abstraction, Interconnected Systems 
		\end{keyword}
		
	\end{frontmatter}
	
	\section{Introduction}
	Model verification and control is an important concept of formal methods, where models are used to represent part of the system that is formalized by a set of properties. It stands for the ability to prove that some properties hold true for a model based on the assumptions of the system and the properties themselves. Model verification and control also serve as powerful tools for validating the correctness and performance of the system. It is a way to ensure that the system meets its specifications when designing a specific control software, see \cite{girard2007approximation,pola2008approximately,tabuada2009verification,julius2009approximate,zamani2011symbolic,hashimoto2019symbolic}. These and other contributions provide a systematic footing for model verification and control of various classes of dynamical systems, including linear, stochastic, and nonlinear systems. Model verification is done by establishing exact or approximate (bi)simulation relations between an original system and its corresponding abstraction, see \cite{girard2007approximation,pola2008approximately,tabuada2009verification}. Abstraction-based controller synthesis responds to synthesize problem of controllers, fulfilling spatio-temporal logic specifications, see \cite{meyer2017compositional,saoud2018composition,saoud2021compositional}. These specifications are usually expressed using temporal logic formula or (in)finite strings over automata.
	
	The abstraction construction procedure generally suffers from scalability issues, making the construction of abstractions challenging for large-scale interconnected systems, see \cite{saoud2019compositional} and references therein. To solve this problem, many compositional approaches have been proposed in the literature. In such approaches, one starts from a large-scale system consisting of interconnected subsystems. Then, an abstraction for the large-scale system is constructed from the abstraction of its subsystems. In this context, (\cite{rungger2016compositional}) relied on the notion of a simulation function and a small-gain type condition to provide a compositional framework that constrains the behavior of the bottom-up system and its abstraction. \cite{zamani2017compositional} and \cite{lavaei2022dissipativity} developed compositional frameworks that quantify the joint dissipativity properties of control subsystems and their abstractions. \cite{swikir2019compositional} has studied the problem of designing controllers of interconnected systems with alternating simulation functions and a small-gain type condition. Finally, \cite{saoud2021compositional} proposed a compositional abstraction framework using the concept of approximate composition, which does not rely on the small-gain condition and results in a more general framework. 
	
	However, all the aforementioned approaches make it possible to compositionally construct an abstraction that is related to the original large-scale system by an approximate (alternating) simulation relation, and cannot be directly generalized to the compositional construction of abstractions that are related to the original system by an approximate (alternating) bisimulation relation. The question of compositional construction of approximately bisimilar abstractions has been only explored in (\cite{tazaki2008bisimilar}). Indeed, given a large-scale system consisting of interconnected components, the authors in (\cite{tazaki2008bisimilar}) show that if each subsystem is related to its abstraction by an interconnection compatible approximate bisimulation relation, then the interconnected system is related to the global abstraction by an approximate bisimulation relation. Moreover, their framework is limited to finite abstractions of interconnected linear subsystems. 

 
	This paper proposes, for the first time in the literature, an approach to compositionally construct approximately bisimilar abstractions for nonlinear systems. Indeed, given a large-scale system consisting of interconnected components, we provide conditions under which the concept of approximate (alternating) simulation relation is preserved when going from the subsystems to the large-scale interconnected system. We rely on the notion of approximate composition introduced in \cite{saoud2021compositional}. This notion allows the distance between inputs and outputs of neighboring components to be bounded by a given parameter called the approximate composition parameter. Indeed, we observe that the behavior of interconnected systems tolerating some composition parameter error becomes more conservative (and less deterministic) when the approximate composition parameters become large. In this paper, we first show how to measure the conservatism (in terms of approximate (alternating) simulation relations) of interconnected systems when enlarging their approximate composition parameters for the case when the subsystems are Incrementally input-to-state stable ($ \delta-ISS $). Indeed, we show that if a collection of $ \delta-ISS $ systems tolerates specific approximate composition errors between adjacent components, then it is approximately (alternatingly) bisimilar to any admissible composition. This preliminary result provides a systematic basis for developing a new framework capable of constructing an abstract system with certain compatibility errors related to the original large-scale exact compatible system by an approximate (alternating) bisimulation relation. As the main advantage, this important new result reduces the number of states and control inputs in the abstract system while maintaining the same transition properties as the concrete system, thus minimizing the computational burden for model verification and controller synthesis.
	\section{Preliminaries and problem statement}
	\textbf{Notations:}
	The symbols $\mathbb{N}, \mathbb{N}_{0}$, $\mathbb{R}$, and $\mathbb{R}_{0}^{+}$ denote the set of positive integers, non-negative integers, real, and non-negative real numbers, respectively.
	For any $x_{1}, x_{2}, x_{3} \in X$, the $\operatorname{map} \mathbf{d}_{X}: X \times X \rightarrow \mathbb{R}_{0}^{+}$ is a \textit{pseudometric} if the following conditions hold: (i) $x_{1}=x_{2}$ implies $\mathbf{d}_{X}\left(x_{1}, x_{2}\right)=0 ;$ (ii) $\mathbf{d}_{X}\left(x_{1}, x_{2}\right)=\mathbf{d}_{X}\left(x_{2}, x_{1}\right)$; (iii) $\mathbf{d}_{X}\left(x_{1}, x_{3}\right) \leq \mathbf{d}_{X}\left(x_{1}, x_{2}\right)+\mathbf{d}_{X}\left(x_{2}, x_{3}\right)$. We identify a relation $\mathcal{R} \subseteq A \times B$ defined by $b \in \mathcal{R}(a)$ if and only if $(a, b) \in \mathcal{R}$. Given a relation $ \mathcal{R}\subseteq A \times B $, $ \mathcal{R}^{-1} $ denotes the inverse relation of $ \mathcal{R} $, i.e. $ \mathcal{R}^{-1}=\left\lbrace (b,a)\in B\times A \mid \quad (a,b) \in R \right\rbrace  $. For $x \in \mathbb{R}^n$,  $\|x\|$ denotes its infinity norm. The null vector of dimension $N \in \mathbb{N}_{0}$ is denoted by $\mathbf{0}_{N}:=(0, \ldots, 0)$. The identity map is denoted by $ \mathrm{id}(s)=s $. For a discrete-time signal, $\mathbf{x}:\mathbb{N}_0 \rightarrow X \subseteq \mathbb{R}^n$, $\|\mathbf{x}\|_{k}= \sup _{j=0,1, \ldots, k} \|\mathbf{x}(j)\|$. The function $\alpha: \mathbb{R}_{\geq 0} \rightarrow \mathbb{R}_{\geq 0}$ is of class $\mathcal{K}$ if it is continuous, $\alpha(0)=0$, and strictly increasing. If $\alpha \in \mathcal{K}$ is unbounded, it is of class $\mathcal{K}_{\infty}$. A function $\sigma: \mathbb{R}_{\geq 0} \rightarrow \mathbb{R}_{\geq 0}$ is of class $\mathcal{L}$ if it is continuous, strictly decreasing, and $\lim _{t \rightarrow \infty} \sigma(t)=0$. A function $\beta: \mathbb{R}_{\geq 0} \times \mathbb{R}_{\geq 0} \rightarrow \mathbb{R}_{\geq 0}$ is of class $\mathcal{K} \mathcal{L}$ if it is class $\mathcal{K}$ in its first argument and class $\mathcal{L}$ in its second argument. For a $ n\times m $ matrix $ A $, $ |A| $ stands for its infinity norm.

	
	%
	
	\subsection{Transition system}
	
	First, we define the \textit{transition systems} adopted from \cite{tabuada2009verification}, which allows us to represent concrete dynamical systems and their abstractions in a unified way.
	\begin{definition}\label{DefTransitionSys}
		A transition system is a tuple $ S=\left(X,X^{0}, \right. $
		$\left. U^{\textit{ext}}, U^{\textit{int}}, \Delta, Y, H\right) $, where $X$ is the set of states (possibly infinite), $X^{0} \subseteq X$ is the set of initial states, $U^{ext}$ and $U^{\textit{int}}$ is the set of external and internal inputs (possibly infinite), respectively, $\Delta \subseteq X \times U^{ext} \times U^{int} \times X$ is the transition relation, $Y$ is the set of outputs, and $H: X \rightarrow Y$ is the output map.
	\end{definition}
	The set of admissible inputs for $x\in X$ is denoted by {\small $U_{S}(x):=\left\{\left({u}^{ext}, {u}^{int}\right) \in U^{ext} \times U^{int} \mid \Delta\left(x, u^{ext}, {u}^{int}\right) \neq \emptyset\right\}$}. Denote by $x^{\prime} \in \Delta\left(x, u^{{ext }}, {u}^{int}\right)$ as an alternative representation for a transition $\left(x, {u}^{ext},{u}^{int}, x^{\prime}\right) \in \Delta$, where state $x^{\prime}$ is called a $\left({u}^{ext}, {u}^{int}\right)$-successor (or simply successor) of state $x$, for some input $\left({u}^{ext}, {u}^{int}\right) \in U^{ext} \times U^{int}$.   A transition system $S$ is said to be:
	\begin{itemize}
		\item[$\bullet$] \textbf{\textit{pseudometric}}, if the state set $X$, input sets $U^{i}, i \in\{ext, int\}$ and the output set $Y$ are equipped with pseudometrics $\mathbf{d}_X:X\times X\rightarrow \mathbb{R}_0^+$, $\mathbf{d}_{U^{i}}: U^{i} \times U^{i} \rightarrow \mathbb{R}_{0}^{+}$ and $\mathbf{d}_Y: Y \times Y \rightarrow \mathbb{R}_{0}^{+}$, respectively;
		\item[$\bullet$] \textbf{\textit{finite}}, if $X$, $ U^{\textit{int}} $, and $ U^{\textit{ext}} $ are finite sets;
		\item[$\bullet$] \textbf{\textit{deterministic}}, if there exists at most one $\left(u^{\textit{ext}}, u^{\textit{int}}\right)$-successor of $x$, for any $x \in X$ and $\left(u^{\textit{ext}}, u^{\textit{int}}\right) \in$ $U^{\textit{ext}} \times U^{\textit{int}}$.
	\end{itemize}
	For a deterministic transition system $S=\left(X, X^{0}, \right. $  $\left.U^{\textit{ext}}, U^{\textit{int}}, \Delta, Y, H\right)$, the notation $\mathbf{x}\left(k,x, \mathbf{u}^{int}, \mathbf{u}^{ext}\right)$ represents the state reached at $k^{th}$ transition from an initial state $x \in X^0 $ under input signals $\mathbf{u}^{int}: \mathbb{N}_0\rightarrow U^{int}$ and $\mathbf{u}^{ext}: \mathbb{N}_0\rightarrow U^{ext} $.
	%
	
	\subsection{Approximate (alternating) bisimulation relations}
	In the following, we introduce a novel notion of approximate (alternating) bisimulation relations, allowing us to relate two transition systems.\\ 
	For two transition systems $S_{1}=(X_{1}, X_{1}^{0}, U_{1}^{ext}, U_{1}^{int}, \Delta_{1}$, $Y_{1}, H_{1})$ and $S_{2}=(X_{2}, X_{2}^{0}, U_{2}^{ext}, U_{2}^{int}, \Delta_{2}, Y_{2}, H_{2})$ such that $Y_{1}$ and $Y_{2}$ are subsets of the same pseudometric space $Y$ equipped with a pseudometric $\mathbf{d}$ and $U_{j}^{ext}$ (respectively $\left.U_{j}^{\mathrm{int}}\right), j \in\{1,2\}$, are subsets of the same pseudometric space $U^{ext}$ (respectively $U^{int}$ ) equipped with a pseudometric $\mathbf{d}_{{u}^{ext}}$ (respectively $\mathbf{d}_{{u}^{int}}$), we introduce the following relations.
	\begin{definition}\label{DefinitionBisimulation}
		For $\varepsilon, \mu \geq 0$, $S_{2}$ is said to be $(\varepsilon, \mu)$-approximately simulated by $S_{1}$, if there exists a relation $\mathcal{R} \subseteq X_{1} \times X_{2}$ satisfying,
		\begin{itemize}
			\item[(i)]$\forall x_{1}^{0} \in X_{1}^{0}, \exists x_{2}^{0} \in X_{2}^{0}$ such that $\left(x_{1}^{0}, x_{2}^{0}\right) \in \mathcal{R}$; 
			\item[(ii)] $\forall\left(x_{1}, x_{2}\right) \in \mathcal{R}, \mathbf{d}\left(H_{1}\left(x_{1}\right), H_{2}\left(x_{2}\right)\right) \leq \varepsilon$;
			\item[(iii)] $\forall\left(x_{1}, x_{2}\right) \in \mathcal{R}, \forall\left({u}_{1}^{ext}, u_{1}^{\mathrm{int}}\right) \in U_{S_{1}}(x), $  $ \forall x_{1}^{\prime} \in \Delta_{1}\left(x_{1}, {u}_{1}^{ext}, u_{1}^{\mathrm{int}}\right), \exists\left({u}_{2}^{ext}, u_{2}^{\mathrm{int}}\right) \in U_{S_{2}}\left(x_{2}\right)$ with $$\max \left(\mathbf{d}_{{u}^{ext}}\left({u}_{1}^{ext}, {u}_{2}^{ext}\right), \mathbf{d}_{{u}^{int}}\left({u}_{1}^{int}, {u}_{2}^{int}\right)\right) \leq \mu$$ and $\exists x_{2}^{\prime} \in \Delta_{2}\left(x_{2}, {u}_{2}^{ext}, {u}_{2}^{int}\right)$ satisfying $\left(x_{1}^{\prime}, x_{2}^{\prime}\right) \in \mathcal{R}$. 
		\end{itemize}
		Moreover, $ S_{2} $ is said to be $ (\varepsilon,\mu) $-approximately bisimilar to $ S_{1} $, if $ S_{1} $ is $ (\varepsilon,\mu) $-approximately simulated by $ S_{2} $, and $ S_{2} $ is $ (\varepsilon,\mu) $-approximately simulated by $ S_{1} $. Simulation and bisimulation relations are denoted respectively by,  $ S_{2} \preccurlyeq^{\varepsilon, \mu} S_{1} $ and $ S_{2} \approx^{\varepsilon, \mu} S_{1} $.
	\end{definition}
	
		For verification problems, approximate (bi)simulation relations are mainly used. The concept of approximate alternating (bi)simulation relations introduced in \cite{tabuada2009verification} are more appropriate if the goal is to synthesize controllers.
	
	\begin{definition}\label{DefinitionAlternatingBisimulation}
		For $\varepsilon, \mu \geq 0$, $S_{2}$ is said to be $(\varepsilon, \mu)$-approximately alternatingly simulated by $S_{1}$, if there exists a relation $\mathcal{R} \subseteq X_{1} \times X_{2}$ satisfying,
		\begin{itemize} \raggedright
			\item[$(\mathrm{i})$]$\forall x_{2}^{0} \in X_{2}^{0}, \exists x_{1}^{0} \in X_{1}^{0}$ such that $\left(x_{1}^{0}, x_{2}^{0}\right) \in \mathcal{R}$;
			\item[$(\mathrm{ii})$] $\forall\left(x_{1}, x_{2}\right) \in \mathcal{R}, \mathbf{d}\left(H_{1}\left(x_{1}\right), H_{2}\left(x_{2}\right)\right) \leq \varepsilon$;
			\item[$(\mathrm{iii})$] $\forall\left(x_{1}, x_{2}\right) \in \mathcal{R}$,  $ \forall\left({u}_{2}^{ext}, {u}_{2}^{int}\right) \in U_{S_{2}}\left(x_{2}\right)$,  $ \exists\left({u}_{1}^{ext}, {u}_{1}^{int}\right) \in U_{S_{1}}\left(x_{1}\right) $   with  $$ \max \left(\mathbf{d}_{{u}^{ext}}\left({u}_{1}^{ext}, {u}_{2}^{ext}\right), \mathbf{d}_{{u}^{int}}\left({u}_{1}^{int}, {u}_{2}^{int}\right)\right) \leq 
			\mu $$ such that $ \forall x_{1}^{\prime} \in \Delta_{1}\left(x_{1}, {u}_{1}^{ext}, {u}_{1}^{int}\right),$ $ \exists x_{2}^{\prime} \in \Delta_{2}\left(x_{2}, {u}_{2}^{ext}, {u}_{2}^{int}\right) $ satisfying $ \left(x_{1}^{\prime}, x_{2}^{\prime}\right) \in \mathcal{R}$ .
		\end{itemize}
		Moreover, $ S_{2} $ is said to be $ (\varepsilon,\mu) $-approximately alternatingly bisimilar to $ S_{1} $, if $ S_{1} $ is $ (\varepsilon,\mu) $-approximately alternatingly simulated by $ S_{2} $, and $ S_{2} $ is $ (\varepsilon,\mu) $-approximately alternatingly simulated by $ S_{1} $. The alternating simulation and alternating bisimulation relations are denoted respectively by $ S_{2} \preccurlyeq_{\mathcal{A}}^{\varepsilon, \mu} S_{1} $ and $ S_{2} \approx_{\mathcal{A}}^{\varepsilon, \mu} S_{1} $.
	\end{definition}


	Contrarily to the concepts of approximate (bi)-simulation relation introduced in \cite{tabuada2009verification} and \cite{girard2007approximation}, the concept of approximate (bi)-simulation relation introduced in \textit{Definition \ref{DefinitionBisimulation}} is more relaxed since it allows a mismatch on the choice of inputs for the transition systems. In particular, when $\mu=0$ and $\mathbf{d}_{\mathbf{u}^{int}}$ is metric, the relation proposed in \textit{Definition \ref{DefinitionBisimulation}} reduces to the notion of approximate bisimulation introduced in \cite{girard2007approximation}, and when $\mu=\infty$, it covers the approximate bisimulation relation given in \cite{tabuada2009verification}. 
	Furthermore, the concept of approximate alternating bisimulation of \textit{Definition \ref{DefinitionAlternatingBisimulation}} includes the one in \cite{pola2009symbolic} by taking $ \mu=\infty $.  
	
	To gather all the ingredients to conduct our main results, the following two propositions are needed. These properties are mainly showing the ordering and the transitivity properties of the introduced relationships. 
	
	\begin{prop}
	\label{prop1}
		Given three pseudometric transition systems $S_{1}, S_{2}$ and $S_{3}$. For any $\mu, \mu^{\prime} \geq 0$ and $\varepsilon, \varepsilon^{\prime} \geq 0$. The following statements hold:
		\begin{itemize}
		\item if $S_{1} \preccurlyeq^{\varepsilon, \mu} S_{2}$ and $S_{2} \preccurlyeq^{\varepsilon^{\prime}, \mu^{\prime}} S_{3}$, then $S_{1} \preccurlyeq^{\varepsilon+\varepsilon^{\prime}, \mu+\mu^{\prime}} S_{3}$
		\item if $S_{1} \preccurlyeq^{\varepsilon, \mu}_{\mathcal{A}} S_{2}$ and $S_{2} \preccurlyeq^{\varepsilon^{\prime}, \mu^{\prime}}_{\mathcal{A}} S_{3}$, then $S_{1} \preccurlyeq^{\varepsilon+\varepsilon^{\prime}, \mu+\mu^{\prime}}_{\mathcal{A}} S_{3}$.
		\end{itemize}
	\end{prop}
	\begin{pf}
		See the Appendix \ref{A1}.$\square$
	\end{pf}
	\begin{prop}
	\label{prop2}
		Given two pseudometric transition systems $S_{1}$ and $S_{2}$. For any $\mu^{\prime} \geq$ $\mu \geq 0$ and $\varepsilon^{\prime} \geq \varepsilon \geq 0$. The following statement holds:
		\begin{itemize}
		\item if $S_{1} \preccurlyeq^{\varepsilon, \mu} S_{2}$ then $S_{1} \preccurlyeq^{\varepsilon^{\prime}, \mu^{\prime}} S_{2}$
		\item if $S_{1} \preccurlyeq_{\mathcal{A}}^{\varepsilon, \mu} S_{2}$ then $S_{1} \preccurlyeq_{\mathcal{A}}^{\varepsilon^{\prime}, \mu^{\prime}} S_{2}$.
		\end{itemize}
	\end{prop}
	\begin{pf}
		See the Appendix \ref{A2}.$\square$
	\end{pf}
	
	\section{Incremental input-to-state Stability for transition systems}
	
	In the following, we introduce the concept of global incremental input-to-state stability $ (\delta-ISS) $ for transition systems.
	\begin{definition} \label{DefD-ISS}
		Consider a deterministic and pseudometric transition system $ S=(X, X^{0}, U^{\textit{ext}}, U^{\textit{int}},\Delta, Y, H) $. The transition system $S$ is said to be globally incrementally Input-to-State Stable ($\delta$-ISS) if there exists a function $\beta$ of class $\mathcal{K} \mathcal{L}$ and a function $\gamma$ of class $\mathcal{K}$ such that,  for any initial states $x_{1}, x_{2} \in X $, for any input signals $\mathbf{u}_{1}^{int}, \mathbf{u}_{2}^{int}: \mathbb{N}_0\rightarrow U^{int}$, $\mathbf{u}_{1}^{ext}, \mathbf{u}_{2}^{ext} : \mathbb{N}_0\rightarrow U^{ext} $, the following inequality holds:
		\begin{align}\label{D-ISS}
			&\resizebox{1\hsize}{!}{$ \mathbf{d}_X(\mathbf{x}\left(k,x_{1}, \mathbf{u}_{1}^{int}, \mathbf{u}_{1}^{ext}\right),\mathbf{x}\left(k,x_{2},\mathbf{u}_{2}^{int}, \mathbf{u}_{2}^{ext}\right)) \leq\beta\left(\mathbf{d}_X(x_{1},x_{2}), k\right) $} \nonumber \\
			&+\hspace{-0.2em}\gamma^{int}\left( \left\|\mathbf{u}_{1}^{int}\hspace{-0.2em}-\hspace{-0.2em}\mathbf{u}_{2}^{int}\right\|_{k-1}\right)\hspace{-0.2em}+ \hspace{-0.2em}\gamma^{ext}\left( \left \|\mathbf{u}_{1}^{ext}\hspace{-0.2em}-\hspace{-0.2em}\mathbf{u}_{2}^{ext}\right\|_{k-1}\right)
		\end{align}
		for all $k \in \mathbb{N}_0$, such that $\mathbf{x}\left(k,x_{1}, \mathbf{u}_{1}^{int}, \mathbf{u}_{1}^{ext}\right) \in X$ and $\mathbf{x}\left(k,x_{2},\mathbf{u}_{2}^{int}, \mathbf{u}_{2}^{ext}\right)\in X $, and where the second and third term of the sum in the right-hand side of \eqref{D-ISS} is taken equal to $ 0 $ for $k=0$.
	\end{definition}
	
	In the rest of the section, we show how to construct the maps $ \beta $ and $ \gamma $ characterizing the $\delta$-ISS properties in \eqref{D-ISS} for discrete-time control systems $(\Sigma_{nl})$ defined as below:
	\begin{align}\label{dt_sys}
		(\Sigma_{nl}):  \mathbf{x}(k+1)=f(\x(k),\uu^{ext}(k),\uu^{int}(k)),\ k\in\mathbb{N}_0,
	\end{align}
	where $\x(k)\in \mathcal{X}$, $\uu^{ext}(k)\in \mathcal{U}^{ext}$, and $\uu^{int}(k)\in \mathcal{U}^{int}$ are state, external and internal inputs, respectively.
	The  discrete-time control system $\Sigma_{nl}$ can be represented as a transition system $ S=\left(X, X^{0}, U^{\textit{ext}}, U^{\textit{int}},\Delta, Y, H\right) $ with $X^{0}=X=\mathcal{X}$, $U^{ext}=\mathcal{U}^{ext}$, $U^{int}=\mathcal{U}^{int}$ the transition $(x,u^{ext},u^{int},x')\in\Delta$ iff $x'=f(x,u^{ext},u^{int})$, for $x,x'\in X$, $u^{ext}\in U^{ext}$ and $u^{int}\in U^{int}$, $Y=X$, and $H(x)=x$. In the rest of the paper, the discrete-time dynamical system $\Sigma_{nl}$ and its transition system's representation $S$ can be used interchangeably.\\
	
\subsection{$ \delta- ISS $ for  discrete-time linear systems}
	Consider a linear  discrete-time system:
	\begin{align}\label{LinearSys}
		(\Sigma_l):\begin{array}{c}
			\x(k+1)=A \x(k)+B \uu^{ext}(k)+D \uu^{int}(k), 
		\end{array}
	\end{align}
	where $ A \in \mathbb{R}^{n \times n}, B \in \mathbb{R}^{n \times m}$, $C \in \mathbb{R}^{q \times n}$ and $D \in \mathbb{R}^{n \times p}$, $ \x, \uu^{ext} $, and $ \uu^{int} $ denote the state signal, the external input signal and the internal input signal, respectively. 
	
	The following result provides conditions for the system $\Sigma_l$ in \eqref{LinearSys} to be $ \delta-ISS $. 
	
	\begin{thm}\label{THM1}
		Consider a system $\Sigma_l$ as in \eqref{LinearSys}. If all the eigen values of the matrix $ A $ are inside the unit disk, then $\Sigma_l$ is $ \delta-ISS $ with functions $ \beta $, $ \gamma^{ext} $, and $\gamma^{int}$ defined, for $(r,k) \in \mathbb{R}_{0}^+ \times \mathbb{N}_{0}$, by:
		\begin{align*}
			&\beta(r, k)=|A^{k}| r, \quad \gamma^{ext}(r) = \frac{|B| r}{1-|A|},\quad \gamma^{int}(r)=\frac{|D| r}{1-|A|}.
		\end{align*}
	\end{thm}
	\begin{pf}
		See the Appendix \ref{A3}.$\square$
  \end{pf}
	\subsection{$ \delta- ISS $ for Lipschitz nonlinear systems}
	Consider the  discrete-time nonlinear system \eqref{dt_sys} satisfies the following Lipschitz continuity assumption:
	\begin{assum}
	\label{assum:4}
		There exist constants $L^{x}, L^{u}, L^{w}\in \mathbb{R}^{+} $ such that:
		\begin{align}\label{Eq28}
			\|f&(x^{a}, {u}^{ext, a},u^{int,a})\hspace{-0.2em}-\hspace{-0.2em}f(x^{b}, u^{ext,b},u^{int,b})\| \hspace{-0.2em}\leq\hspace{-0.2em} L^{x}\|x^{a}\hspace{-0.2em}-\hspace{-0.2em}x^{b}\| \nonumber \\
			&+ \hspace{-0.2em}L^{u^{ext}}\|u^{ext,a}\hspace{-0.2em}-\hspace{-0.2em}u^{ext, b}\| 
        \hspace{-0.2em}+\hspace{-0.2em} L^{u^{int}}\left\|u^{int,a}\hspace{-0.2em}-\hspace{-0.2em}u^{int,b}\right\|
		\end{align}
		$\forall x^{a}, x^{b} \hspace{-0.2em}\in\hspace{-0.2em} X, \forall u^{int,a}, u^{int,b} \hspace{-0.2em}\in \hspace{-0.2em}U^{int} \textrm{and } \forall u^{ext,a}, u^{ext,b}\hspace{-0.2em} \in\hspace{-0.2em} U^{ext}.$
	\end{assum}
	The following  result is adapted from \textit{Theorem 1} in \cite{bayer2013discrete}. 
	
	
	\begin{thm}\label{THM2}
		Given a system $\Sigma_{nl}$ in \eqref{dt_sys} satisfying Assumption \ref{assum:4}. If the constant $ L^{x} $ satisfies $ L^{x} <1$, then $\Sigma_{nl}$ is $ \delta-ISS $ with functions $ \beta $, $ \gamma^{ext} $, and $\gamma^{int}$ defined, for $(r,k) \in \mathbb{R}_{0}^+ \times \mathbb{N}_{0}$, by: 
		\begin{align*}
			&\beta(r, k)=(L^{x})^{k} r, \gamma^{ext}(r) = \frac{(L^{u^{ext}}) r}{1-L^{x}}, \gamma^{int}(r)=\frac{(L^{u^{int}}) r}{1-L^{x}}.
		\end{align*}
  \end{thm}
  
  \begin{pf}
		See the Appendix \ref{A4}.$\square$
  \end{pf}


	\section{Compositional bisimilar abstractions for interconnected systems}
	
	In this section, we consider networks of interconnected transition systems. We also state our main result, by providing conditions to preserve approximate (alternating) bisimulation relations from the subsystems to the global interconnected system.

	\subsection{Interconnected system}
	
	An interconnected system is composed of a collection of $ N \in \mathbb{N} $ transition systems $\left\{S_{i}\right\}_{i \in I}$,  a set of vertices $ I = \left\lbrace 1,\dots, N \right\rbrace  $ and a binary connectivity relation $ \mathcal{I} \subseteq I \times I $ where each vertex $ i \in I $ is labelled with the system $ S_{i} $. For $ i \in I $, we define $ \mathcal{N}(i) = \left\lbrace j \in I | (j, i) \in \mathcal{I} \right\rbrace $ as the set of neighbouring components from where the incoming edges come. The $ i^{th} $ subsystem is described by $S_{i}=\left(X_{i}, X_{i}^{0}, U_{i}^{\textit{ext}}, U_{i}^{\mathrm{int}}, \Delta_{i}, Y_{i}, H_{i}\right)$, where $H_{i}$ is an identity map $ H_{i}(x)=x $. 
\begin{definition}\label{DefComposition}
		Given a collection of transition systems $\left\{S_{i}\right\}_{i \in I}$, where $S_{i}=\left(X_{i}, X_{i}^{0}, U_{i}^{ext}, U_{i}^{int}, \Delta_{i}, Y_{i}, H_{i}\right)$ such that for all $i \in I, \prod_{j \in \mathcal{N}(i)} Y_{j}$ and $U_{i}^{\mathrm{int}}$ are subsets of the same pseudometric space equipped with the following pseudometric:
		\begin{align*}
			\text{for }u_{i}^{l {,int }}&\hspace{-0.2em}=\hspace{-0.2em}\left(y_{j_{1}}^{l}, \ldots, y_{j_{k}}^{l}\right)\hspace{-0.2em}, l \hspace{-0.2em}\in\hspace{-0.2em}\{1,2\},\text{ with }\mathcal{N}(i)\hspace{-0.2em}=\hspace{-0.2em}\left\{j_{1}, \ldots, j_{k}\right\}\hspace{-0.2em},\\ &\mathbf{d}_{U_{i}^{\mathrm{int}}}\left(u_{i}^{1, \mathrm{int}}, u_{i}^{2, { int }}\right)=\max _{j \in \mathcal{N}(i)}\left\{\mathbf{d}_{Y_{j}}\left(y_{j}^{1}, y_{j}^{2}\right)\right\}.
		\end{align*}
		Let $M:=\left(\mu_{1}, \ldots, \mu_{N}\right) \in\left(\mathbb{R}_{0}^{+}\right)^{N}$. We say that $\left\{S_{i}\right\}_{i \in I}$ is compatible for $M$-approximate composition with respect to $\mathcal{I}$, if for each $i \in I$ and for each $\prod_{j \in \mathcal{N}(i)}\left\{y_{j}\right\} \in \prod_{j \in \mathcal{N}(i)} Y^{j}$, where the term $\prod_{j \in \mathcal{N}(i)}\left\{y_{j}\right\}$ can be formally defined as $\prod_{j \in \mathcal{N}(i)}\left\{y_{j}\right\}=\left(y_{j_{1}}, y_{j_{2}}, \ldots, y_{j_{p}}\right)$ with $\mathcal{N}(i)=\left\{j_{1}, j_{2}, \ldots, j_{p}\right\}$, there exists $u_{i}^{int} \in U_{i}^{int}$ such that $\mathbf{d}_{U_{i}^{\mathrm{int}}}\left(u_{i}^{\mathrm{int}}, \prod_{j \in \mathcal{N}(i)}\left\{y_{j}\right\}\right) \leq \mu_{i}$. We denote $M$-approximate composed system by $\left\langle S_{i}\right\rangle_{i \in I}^{M, \mathcal{I}}$ and is given by the tuple $\left\langle S_{i}\right\rangle_{i \in I}^{M, \mathcal{I}}=\left(X, X^{0}, U^{ext}, \Delta_{M}, Y, H\right)$, where:
		\begin{itemize}\raggedright
			\item $X=\prod_{i \in I} X_{i}$; $X^{0}=\prod_{i \in I} X_{i}^{0}$; $U^{ext}=\prod_{i \in I} U_{i}^{ext}$; $Y=\prod_{i \in I} Y_{i}$;
			\item $H(x)=H\left(x_{1}, \ldots, x_{N}\right)=\left(H_{1}\left(x_{1}\right), \ldots, H_{N}\left(x_{N}\right)\right)=(x_{1}, \ldots, x_{N})$
			\item for $x=\left(x_{1}, \ldots, x_{N}\right), x^{\prime}=\left(x_{1}^{\prime}, \ldots, x_{N}^{\prime}\right)$ and ${u}^{ext}=\left({u}_{1}^{ext}, \ldots, u_{N}^{ext}\right), x^{\prime} \in \Delta_{M}\left(x, {u}^{ext}\right)$ if and only if for all $i \in I$, and for all $\prod_{j \in \mathcal{N}(i)}\left\{y_{j}\right\}=\prod_{j \in \mathcal{N}(i)}\left\{H_{j}\left(x_{j}\right)\right\} \in \prod_{j \in \mathcal{N}(i)} Y_{j}$, there exists $u_{i}^{int} \in U_{i}^{int}$ with $\mathbf{d}_{U_{i}^{int}}\left(u_{i}^{int}, \prod_{j \in \mathcal{N}(i)}\left\{y_{j}\right\}\right) \leq \mu_{i},\left(u_{i}^{ext}, u_{i}^{int}\right) \in U_{S_{i}}\left(x_{i}\right)$ and $x_{i}^{\prime} \in \Delta_{i}\left(x_{i}, u_{i}^{ext}, u_{i}^{int}\right)$.
		\end{itemize}
	\end{definition}

\begin{example}
    An example of the interconnection of three transition systems $S_1, S_2$ and $S_3$, is presented in Fig. \ref{fig:interconnectedsys}(a).  The connectivity relation is defined by $\mathcal{I}= \{(1,2), (2,3), (3,1)\}$, and $\mathcal{N}(1)=3, \; \mathcal{N}(2)=1, \; \mathcal{N}(3)=2$.  Fig. \ref{fig:interconnectedsys}(b) illustrates the case of a composition with an approximate composition parameter $M=(\mu_1,\mu_2,\mu_3)$.
    
 \begin{figure}
\begin{subfigure}{0.5\textwidth}
\flushleft
 \vspace{-0.4cm}
\includegraphics[width=1\linewidth]{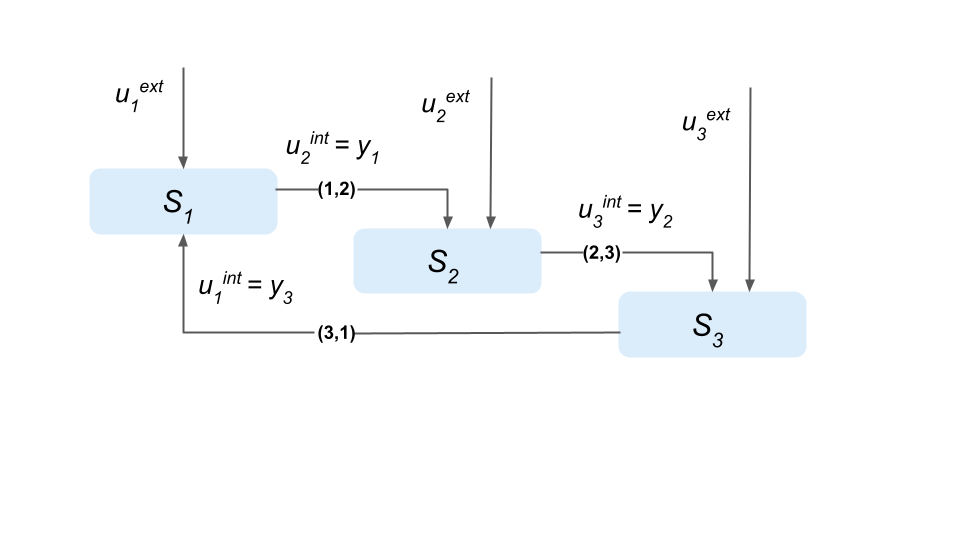}\\
\vspace{-1.5cm}
\subcaption{Example of exact composition.}
\end{subfigure}
\begin{subfigure}{0.5\textwidth}
\includegraphics[width=1\linewidth]{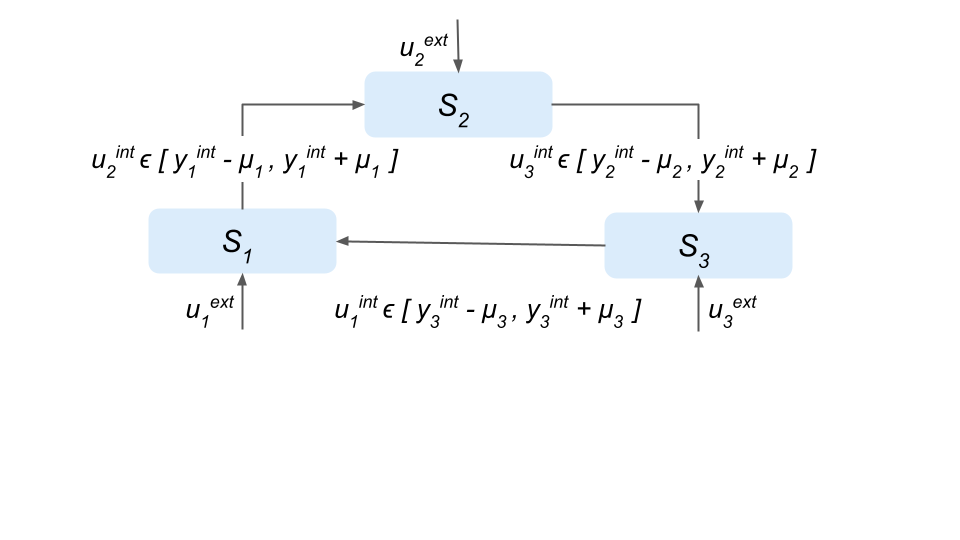}
\vspace{-2.1cm}
\subcaption{Example of composition with an approximate composition parameter $M=(\mu_1,\mu_2,\mu_3)$.}
\end{subfigure}
\caption{Interconnection of three systems with different approximate composition parameters.}
\label{fig:interconnectedsys}
\end{figure}
\end{example}
 
  We equip the composed output space with the metric:
	\begin{align}\label{Eq10}
		for \; y^{j} \in Y { \; with \; } y^{j}=(y_{1}^{j}, \ldots, y_{N}^{j}), j \in\{1,2\}, \nonumber\\
		\mathbf{d}\left(y^{1}, y^{2}\right)=\max _{i \in I}\left\{\mathbf{d}_{Y_{i}}\left(y_{i}^{1}, y_{i}^{2}\right)\right\}
	\end{align}
	Similarly, we equip the composed input and the state spaces with the pseudometric:
	\begin{align}\label{Eq11}
	&\resizebox{0.95\hsize}{!}{$for \; u^{j,ext} \in U^{ext} { \; with \; } u^{j,ext}=(u_{1}^{j,ext}, \ldots, u_{N}^{j,ext}), j \in\{1,2\},$} \nonumber\\ 
 &\mathbf{d}_{U^{ext}}(u^{1,ext}, u^{2,ext})=\max _{i \in I}\left\{\mathbf{d}_{U_{i}^{ext}}\left(u_{i}^{1,ext}, u_{i}^{2,,ext}\right)\right\} ,
	\end{align}
\begin{align}
    &\resizebox{0.95\hsize}{!}{$for \; u^{j,int} \in U^{int} { \; with \; } u^{j,int}=(u_{1}^{j,int}, \ldots, u_{N}^{j,int}), j \in\{1,2\},$} \nonumber\\ &\mathbf{d}_{U^{int}}(u^{1,int}, u^{2,int})=\max _{i \in I}\left\{\mathbf{d}_{U_{i}^{int}}\left(u_{i}^{1,int}, u_{i}^{2,,int}\right)\right\} .
\end{align}
	The behavior of interconnected systems tolerating some composition parameter error becomes more conservative (and less deterministic) when the approximate composition parameters become large. The following result shows that under the $ \delta $-ISS property, we can measure the conservatism of the approximate composition when increasing the approximate composition parameter.


	\begin{thm}\label{THM6}
		Consider a collection of transition systems $\left\{S_{i}\right\}_{i \in I}$ and $\bar{M}=(\bar{\mu}_{1}, \ldots, \bar{\mu}_{N} \in(\mathbb{R}_{0}^{+})^{N}$. If each subsystem of $\left\{S_{i}\right\}_{i \in I}$ is $ \delta $-ISS and  $\left\{S_{i}\right\}_{i \in I}$ is compatible for $\bar{M}$-approximate composition with respect to $\mathcal{I}$, then it is also compatible for $M$-approximate composition with respect to $\mathcal{I}$, for any $M=\left(\mu_{1}, \ldots, \mu_{N}\right) \in(\mathbb{R}_{0}^{+})^{N}$ such that $\bar{M} \geq M$ (i.e., $\left.\bar{\mu}_{i} \geq \mu_{i}, i \in I\right)$. Moreover, for any $\varepsilon \geq 0$ such that 
		\begin{equation}
		\label{eqn:thm6}
		    \beta_{i}(\varepsilon,1) + \gamma^{int}_{i}(\varepsilon+\bar{\mu}_i-\mu_i) \leq \varepsilon,~~  \forall i \in I
		\end{equation}
		 the relation $\mathcal{R}= \{ \left(x, x^{\prime}\right)  \in X \times X \: | \: \mathbf{d}(H(x),H(x^{\prime}))\leq \varepsilon \}$ is a $ (\varepsilon,0) $-approximate bisimulation relation between  $S_{\bar{M}}=\left\langle S_{i}\right\rangle_{i \in I}^{\bar{M}, \mathcal{I}}$ and $ S_{M}=\langle S^{i}\rangle_{i \in I}^{M, \mathcal{I}} $.
	\end{thm}
	\begin{pf}
	 It was shown in \cite{saoud2021compositional} that $ S_{M} \preccurlyeq^{0, 0} S_{\bar{M}} $ is verified for $ \bar{M}\geqslant M$, which implies from Proposition \ref{prop2} that $S_{M} \preccurlyeq^{\varepsilon, 0} S_{\bar{M}}$ . The rest of this proof will focus on showing that the condition \eqref{eqn:thm6} guarantees the symmetrical version $ S_{\bar{M}} \preccurlyeq^{\varepsilon, 0} S_{M} $. \\
		The first and second conditions in \textit{Definition \ref{DefinitionBisimulation}} are directly satisfied. (We have the same transition systems, with the same set of initial conditions and the same sets of states, thus, the distance between two elements $x^{a}$ and $x^{b}$, satisfying $(x^{a},x^{b})\in \mathcal{R}$ is bounded by $\varepsilon$). 
		
		Consider $ (x^{a},x^{b})\in \mathcal{R} $, with $ x^{a}=(x_{1}^{a}, \dots, x_{N}^{a}) $ and $ x^{b}=(x_{1}^{b}, \dots, x_{N}^{b}) $, any $ u^{a,ext} \in U_{S_{\bar{M}}}(x^{a}) $ and any $ x^{a\prime} \in \Delta_{\bar{M}}(x^{a},u^{a,ext}) $. Choose $ u^{b,ext}=u^{a,ext} $ and let us show the existence of $x^{b\prime} \in \Delta_{M}(x^{b}, u^{b,ext})$ satisfying $\left(x^{a\prime}, x^{b\prime}\right) \in \mathcal{R}$. Since, $x^{a\prime} \in \Delta_{\bar{M}}(x^{a},u^{a,ext})$, there exists $u_{i}^{a,int} \in U_{i}^{a,int}$ with $\mathbf{d}_{U_{i}^{int}}(u_{i}^{a,int}, \prod_{j \in \mathcal{N}(i)}\{y^a_{j}\}) \leq \bar{\mu}_{i},(u_{i}^{a,ext}, u_{i}^{a,int}) \in U_{S_{i}}(x^a_{i})$ and $x_{i}^{a\prime} \in \Delta^a_{i}(x^a_{i}, u_{i}^{a,ext}, u_{i}^{a,int})$.
		
		Now for $i \in I$, choose $u_{i}^{b,int} \in U_i^{int}$ satisfying $\mathbf{d}_{U_{i}^{int}}(u_{i}^{b,int}, $ $ \prod_{j \in \mathcal{N}(i)}\{y^b_{j}\})\leq \mu_i$ and $\mathbf{d}_{U_{i}^{int}}(u_{i}^{b,int},u_{i}^{a,int}) \leq \varepsilon +\bar{\mu_{i}}-\mu_i$. Such $u_{i}^{b,int}$ always exists using the fact that $\mathbf{d}(\prod_{j \in \mathcal{N}(i)}\left\{y^a_{j}\right\},\prod_{j \in \mathcal{N}(i)}\left\{y^b_{j}\right\}) \leq \varepsilon$ and $\mathbf{d}_{U_{i}^{int}}(u_{i}^{a,int},$ $\prod_{j \in \mathcal{N}(i)}\{y^a_{j}\}) \leq \bar{\mu}_{i}$. Now pick $x^{b\prime} \in \Delta_{\bar{M}}(x^{b},u^{b,ext})$, defined for $i \in I$ as $x_{i}^{b\prime} \in \Delta^b_{i}(x^b_{i}, u_{i}^{b,ext}, u_{i}^{b,int})$. First, we have that for the chosen values of $u_{i}^{b,int}$, $i \in I$, the transition $x^{b\prime} \in \Delta_{\bar{M}}(x^{b},u^{b,ext})$ is well defined for the approximate composed system $S_M$. Let us now show that $(x^{a\prime}, x^{b\prime}) \in \mathcal{R}$. From the definition of the relation $ \mathcal{R} $ and under the assumption that each component satisfies the $ \delta $-ISS property, we have for all $ i\in I  $, 
		\begin{align*}
		    \mathbf{d}(x^{a\prime},x^{b\prime}) &\resizebox{0.8\hsize}{!}{$\leq
			\max _{i \in I}\mathbf{d}_{Y_i}(\Delta_{i}(x_{i}^{a},u_{i}^{a,ext}, u_{i}^{a,int}),\Delta_{i}(x^b_{i},u_{i}^{b,ext},u_{i}^{b,int}))$}, \nonumber\\
			&\leq \max _{i \in I}  \beta_{i}(\varepsilon,1) + \gamma^{int}_{i}(\varepsilon+\bar{\mu_{i}}-\mu_i) \leq \varepsilon,
		\end{align*}
		where the last inequality follows from (\ref{eqn:thm6}). Hence, condition (iii) in \textit{Definition \ref{DefinitionBisimulation}} holds and one obtains $S_{\bar{M}} \preccurlyeq^{\varepsilon, 0} S_{M}$.$\square$
	\end{pf}
	
	\begin{rem}It is worth noticing that the results in \cite{saoud2021compositional} shows a simulation relation $ S_{M} \preccurlyeq^{\varepsilon,0} S_{\bar{M}} $. The previous result shows that under the $ \delta-ISS $ property of each subsystem the symmetrical relation  $ S_{\bar{M}} \preccurlyeq^{\varepsilon,0} S_{M} $ holds and thus $ S_{\bar{M}} \approx ^{\varepsilon,0}  S_{M} $. Indeed, while any trajectory of the system $S_{M}$ is a trajectory of the system $S_{\bar{M}}$, the proposed result shows that under the $\delta-ISS$ property, one can measure the conservatism between $S_{M}$ and $S_{\bar{M}}$, thereby measuring the conservatism of the approximate composition.
	\end{rem}

	\begin{thm}\label{THM9}
		Consider a collection of transition systems $\left\{S_{i}\right\}_{i \in I}$ and $\bar{M}=\left(\bar{\mu}_{1}, \ldots, \bar{\mu}_{N}\right)  \in\left(\mathbb{R}_{0}^{+}\right)^{N}$. If each subsystem of $\left\{S_{i}\right\}_{i \in I}$ is $ \delta $-ISS and  $\left\{S_{i}\right\}_{i \in I}$ is compatible for $\bar{M}$-approximate composition with respect to $\mathcal{I}$, then it is also compatible for $M$-approximate composition with respect to $\mathcal{I}$, for any $M=\left(\mu_{1}, \ldots, \mu_{N}\right) \in\left(\mathbb{R}_{0}^{+}\right)^{N}$ such that $\bar{M} \geq M$ (i.e., $\left.\bar{\mu}_{i} \geq \mu_{i}, i \in I\right)$. Moreover, for any $\varepsilon \geq 0$ such that 
		\begin{equation}
		\label{eqn:thm8}
		    \beta_{i}(\varepsilon,1) + \gamma^{int}_{i}(\varepsilon+\bar{\mu}_i-\mu_i) \leq \varepsilon,~~  \forall i \in I
		\end{equation}
		 the relation $\mathcal{R}= \{ \left(x, x^{\prime}\right)  \in X \times X \: | \: \mathbf{d}(H(x),H(x^{\prime}))\leq \varepsilon \}$ is a $ (\varepsilon,0) $-approximate alternating bisimulation relation between  $S_{\bar{M}}=\left\langle S_{i}\right\rangle_{i \in I}^{\bar{M}, \mathcal{I}}$ and $ S_{M}=\langle S^{i}\rangle_{i \in I}^{M, \mathcal{I}} $.
	\end{thm}
	\begin{pf}
 It was shown in \cite{saoud2021compositional} that $ S_{M} \preccurlyeq_{\mathcal{A}}^{0, 0} S_{\bar{M}} $ is verified for $ \bar{M}\geqslant M$, which implies from Proposition \ref{prop2} that $S_{M} \preccurlyeq_{\mathcal{A}}^{\varepsilon, 0} S_{\bar{M}}$ . The rest of this proof will focus on showing that the condition \eqref{eqn:thm8} guarantees the symmetrical version $ S_{\bar{M}} \preccurlyeq_{\mathcal{A}}^{\varepsilon, 0} S_{M} $. 
		The first and second conditions in \textit{Definition \ref{DefinitionAlternatingBisimulation}} are directly satisfied.
		
		Consider $ (x^{a},x^{b})\in \mathcal{R} $, with $ x^{a}=(x_{1}^{a}, \dots, x_{N}^{a}) $ and $ x^{b}=(x_{1}^{b}, \dots, x_{N}^{b}) $, any $ u^{b,ext} \in U_{S_{\bar{M}}}(x^{b}) $ and choose $ u^{a,ext}=u^{b,ext} $. Let us prove that for any $ x^{a\prime} \in \Delta_{\bar{M}}  (x^{a},u^{a,ext})$, there exists $x^{b\prime} \in \Delta_{M}(x^{b}, u^{b,ext})$ satisfying $\left(x^{a\prime}, x^{b\prime}\right) \in \mathcal{R}$. 
		
		Consider $x^{a\prime} \in \Delta_{\bar{M}}(x^{a},u^{a,ext})$, there exists $u_{i}^{a,int} \in U_{i}^{a,int}$ with $\mathbf{d}_{U_{i}^{int}}(u_{i}^{a,int}, \prod_{j \in \mathcal{N}(i)}\left\{y^a_{j}\right\}) \leq \bar{\mu}_{i},(u_{i}^{a,ext}, u_{i}^{a,int}) \in U_{S_{i}}\left(x^a_{i}\right)$ and $x_{i}^{a\prime} \in \Delta^a_{i}(x^a_{i}, u_{i}^{a,ext}, u_{i}^{a,int})$. 
		
		Now for $i \in I$, choose $u_{i}^{b,int} \in U_i^{int}$ satisfying $\mathbf{d}_{U_{i}^{int}}(u_{i}^{b,int},$ $\prod_{j \in \mathcal{N}(i)}\left\{y^b_{j}\right\})\leq \mu_i$ and $\mathbf{d}_{U_{i}^{int}}(u_{i}^{b,int},$ $u_{i}^{a,int}) \leq \varepsilon + \bar{\mu_{i}} -\mu_i$. Such $u_{i}^{b,int}$ always exists using the fact that $\mathbf{d}(\prod_{j \in \mathcal{N}(i)}\left\{y^a_{j}\right\},\prod_{j \in \mathcal{N}(i)}\left\{y^b_{j}\right\}) \leq \varepsilon$ and $\mathbf{d}_{U_{i}^{int}}\left(u_{i}^{a,int},\right.$ $\left. \prod_{j \in \mathcal{N}(i)}\left\{y^a_{j}\right\}\right) \leq \bar{\mu}_{i}$. Now pick $x^{b\prime} \in \Delta_{\bar{M}}(x^{b},u^{b,ext})$, defined for $i \in I$ as $x_{i}^{b\prime} \in \Delta^b_{i}\left(x^b_{i}, u_{i}^{b,ext}, u_{i}^{b,int}\right)$. First, we have that for the chosen values of $u_{i}^{b,int}$, $i \in I$, the transition $x^{b\prime} \in \Delta_{\bar{M}}(x^{b},u^{b,ext})$ is well defined for the approximate composed system $S_M$. Let us now show that $\left(x^{a\prime}, x^{b\prime}\right) \in \mathcal{R}$. From the definition of the relation $ \mathcal{R} $ and under the assumption that each component satisfies the $ \delta $-ISS property, we have for all $ i\in I  $, 
		\begin{align*}
		    \mathbf{d}(x^{a\prime},x^{b\prime}) & \resizebox{0.8\hsize}{!}{$=
			\max _{i \in I}\mathbf{d}_{Y_i}(\Delta_{i}(x_{i}^{a},u_{i}^{a,ext}, u_{i}^{a,int}),\Delta_{i}(x^b_{i},u_{i}^{b,ext},u_{i}^{b,int})) $}\nonumber\\
			&\leq \max _{i \in I}  \beta_{i}(\varepsilon,1) + \gamma^{int}_{i}(\varepsilon+\bar{\mu_{i}}-\mu_i) \leq \varepsilon,
		\end{align*}
		where the last inequality follows from (\ref{eqn:thm8}). Thus, condition (iii) in \textit{Definition \ref{DefinitionAlternatingBisimulation}} holds, and one obtains $S_{\bar{M}} \preccurlyeq_\mathcal{A}^{\varepsilon, 0} S_{M}$. $\square$
	\end{pf}
	\begin{rem}
		It is worth noticing that we have the alternating simulation relation from $ S_{M} \preccurlyeq_{\mathcal{A}}^{\varepsilon,0} S_{\bar{M}} $ without any stability requirement. The proposed result shows the symmetrical version $ S_{\bar{M}} \preccurlyeq_{\mathcal{A}}^{\varepsilon,0} S_{M}  $ under the  $ \delta-ISS $ property of each subsystem.
	\end{rem}
	
	\subsection{Approximate Bisimilar Composition}
	The compositionality result for approximate bisimulation relation is stated as follows.
	\begin{thm}\label{THM10} Let $\{S_{i}\}_{i \in I}$ and $\{\hat{S}_{i}\}_{i \in I}$ be two collections of transition systems with $S_{i}=(X_{i}, X_{i}^{0}, U_{i}^{ext}, U_{i}^{\mathrm{int}}, \Delta_{i}$, $Y_{i}, H_{i})$ and $\hat{S}_{i}=(\hat{X}_{i}, \hat{X}_{i}^{0}, \hat{U}_{i}^{ext}, \hat{U}_{i}^{int}, \hat{\Delta}_{i}, \hat{Y}_{i}, \hat{H}_{i})$. Consider positive constants $\varepsilon_{i}, \mu_{i},$ for $ i \in I$, with $\varepsilon=\max _{i \in I} \varepsilon_{i}$, $\mu=\max_{i \in I} \mu_{i}$ and consider $M=\left(\delta_{1}, \ldots, \delta_{N}\right) $ and $\hat{M}=(\mu_1+\delta_1+\varepsilon,\ldots, \mu_N+\delta_N+\varepsilon)$ . Let the following hold:
		\begin{itemize}
			\item[$(\mathrm{i})$]  For   all  $i \in I$,  $ S_{i} $ is $ \delta $-ISS and satisfies the following inequality $ \max_{i \in I} \big(\beta_{i}(\varepsilon,1) + \gamma^{int}_{i}(2\varepsilon+\mu_i)  \big)\leq \varepsilon$;
			\item[$(\mathrm{ii})$]  For   all  $i \in I$,  $ S_{i} $ is  $ (\varepsilon_{i},\mu_{i}) $- approximately bisimilar  to  $ \hat{S}_{i} $, and we denote $ S_{i} \approx^{\varepsilon_{i},\mu_{i}} \hat{S}_{i} $;
			\item[$(\mathrm{iii})$] $\left\{S_{i}\right\}_{i \in I}$ are compatible for $M$-approximate composition with respect to $\mathcal{I}$;
			\item[$(\mathrm{iv})$] $\{\hat{S}_{i}\}_{i \in I}$ are compatible for $\hat{M}$-approximate composition with respect to $\mathcal{I}$;
		\end{itemize}
		then, ${S}_{{M}}=\left\langle {S}_{i}\right\rangle_{i \in I}^{{M}, \mathcal{I}}$ is  $(2\varepsilon,\mu)$-approximately bisimilar to $\hat{S}_{\hat{M}}=\langle \hat{S}_{i}\rangle_{i \in I}^{\hat{M}, \mathcal{I}}$.
	\end{thm}
	\begin{pf}
        First, we have from \cite{saoud2021compositional} that $ {S}_{{M}} \preccurlyeq^{\varepsilon,\mu}  \hat{S}_{\hat{{M}}}$. Hence, one gets from Proposition \ref{prop2} that ${S}_{{M}} \preccurlyeq^{2\varepsilon,\mu}  \hat{S}_{\hat{{M}}}$. Let us now show that $\hat{S}_{\hat{M}}\preccurlyeq^{2\varepsilon,\mu} {S}_{{M}}$.
		
		For $i \in I$, let $\mathcal{R}_i$ be the $(\varepsilon_i,\mu_i)$-approximate simulation relation from $\hat{S}_i$ to $S_i$, and let us first show that the relation $\mathcal{R}$ defined by 
		$\mathcal{R}= \{ \left(x, \hat{x}\right)  \in X \times \hat{X} \: | \: (x_i,\hat{x}_i) \in \mathcal{R}_i \}$, with $x=\left(x_{1}, \ldots, x_{N}\right)$ and $\hat{x}=\left(\hat{x}_{1}, \ldots, \hat{x}_{N}\right)$, is an $(\varepsilon,\mu)$-approximate simulation relation from $\hat{S}_{\hat{M}}$ to $S_{\bar{M}}$, with $\bar{M}=(2\mu_1+\delta_1+2\varepsilon,\ldots, 2\mu_N+\delta_N+2\varepsilon)$. 
		
		The first condition is directly satisfied (we suppose that for all $ i\in N $, $ \hat{S}_{i} $ is $(\varepsilon_i,\mu_i)$-approximately similar to $ S_{i} $, thus, for each initial condition in $ \hat{S}_{\hat{M}} $ we can find an initial condition in $ S_{\bar{M}} $).

		Let $(x, \hat{x}) \in \mathcal{R}$ with $x=\left(x_{1}, \ldots, x_{N}\right)$ and $\hat{x}=\left(\hat{x}_{1}, \ldots, \hat{x}_{N}\right)$. Using the definition of the output map for approximate composition, relation \eqref{Eq10} and condition (ii) of \textit{Definition \ref{DefinitionBisimulation}}, we have, 
		\begin{align*}
			&\resizebox{1\hsize}{!}{$ \mathbf{d}(H(x), \hat{H}(\hat{x}))=\mathbf{d}(\left(H_{1}\left(x_{1}\right), \ldots, H_{N}\left(x_{N}\right)\right),(\hat{H}_{1}\left(\hat{x}_{1}\right), \ldots, \hat{H}_{N}\left(\hat{x}_{N}\right))) $}\nonumber \\
			&=\max _{i \in I} \mathbf{d}_{Y_{i}}(H_{i}\left(x_{i}\right), \hat{H}\left(\hat{x}_{i}\right)) \leq \max _{i \in I} \varepsilon_{i}=\varepsilon
		\end{align*}
		where the last inequality follows from the fact that $(x_i,\hat{x}_i) \in \mathcal{R}_i$ for all $i \in I$.
		
		Consider $(x,\hat{x})\in\mathcal{R}$ with $x=(x_1,\ldots,x_N)$ and $\hat{x}=(\hat{x}_1,\ldots,\hat{x}_N)$, any $\hat{u}^{ext} \in \hat{U}_{\hat{S}_{\hat{M}}}(\hat{x})$ with $\hat{u}^{ext}=(\hat{u}_1^{ext},\ldots,\hat{u}_N^{ext})$ and any $\hat{x}' \in \hat{\Delta}_{\hat{M}}(\hat{x},\hat{u})$. Let us prove the existence of $u^{ext} \in U_{S_{\bar{M}}}(x)$ with $\mathbf{d}_{U^{ext}}(u^{ext},\hat{u}^{ext}) \leq \mu$ and ${x}' \in {\Delta}_{\bar{M}}({x},{u})$ satisfying $(x',\hat{x}')\in \mathcal{R}$.
		
	From the definition of the relation $\mathcal R$, we have for all $i\in I$, $(x_i,\hat{x}_i)\in \mathcal{R}_i$, then from the third condition of \textit{Definition \ref{DefinitionBisimulation}}, we have for all $(\hat{u}_i^{ext},\hat{u}_i^{int}) \in \hat{U}_{\hat{S}_i}(\hat{x}_i)$, and for any $\hat{x}_i'\in \hat{\Delta}_i(\hat{x}_i,\hat{u}_i^{ext},\hat{u}_i^{int})$ the existence of $(u_i^{ext},u_i^{int})\in U_{S_i}(x_i)$ with $\mathbf{d}_{U_i^{ext}}(u_i^{ext},\hat{u}_i^{ext})\leq\mu_i$ and $\mathbf{d}_{U_i^{int}}(u_i^{int},\hat{u}_i^{int})\leq\mu_i$ and the existence of ${x}_i'\in {\Delta}_i({x}_i,{u}_i^{ext},{u}_i^{int})$ such that $(x_i',\hat{x}_i')\in \mathcal{R}_i$.
	
	Let us show that the input ${u}^{int}=({u}^{int}_1,\ldots,{u}^{int}_N)$ satisfies the requirement of the $\bar{M}$-approximate composition of the components $\{{S}_i\}_{i\in I}$. The condition $\mathbf{d}_{U_i^{int}}(u_i^{int},\hat{u}_i^{int})\leq \mu_i$ implies that,
\begin{align*}
&\resizebox{1\hsize}{!}{$\mathbf{d}_{U_i^{int}}({u}_i^{int},\prod_{j\in \mathcal{N}(i)}\{{y}_j\}) \leq \mathbf{d}_{U_i^{int}}(\hat{u}_i^{int},u_i^{int})+\mathbf{d}_{U_i^{int}}(\hat{u}_i^{int},\prod_{j\in \mathcal{N}(i)}\{{y}_j\})$} \\
		& \resizebox{1\hsize}{!}{$\leq \mathbf{d}_{U_i^{int}}(\hat{u}_i^{int},u_i^{int})+\mathbf{d}_{U_i^{int}}(\hat{u}_i^{int},\prod_{j\in \mathcal{N}(i)}\{\hat{y}_j\})+\mathbf{d}_{U_i^{int}}(\prod_{j\in \mathcal{N}(i)}\{y_j\},\prod_{j\in \mathcal{N}(i)}\{\hat{y}_j\})$}\\
& \resizebox{1\hsize}{!}{$\leq \mu_i + \mu_i+\delta_i+\varepsilon+ \max\limits_{j\in \mathcal{N}(i)} \varepsilon_j \leq 2\mu_i +\delta_i+\varepsilon+ \max\limits_{j \in I} \varepsilon_j =2\mu_i+\delta_i+2\varepsilon.$}
\end{align*}
	Hence, from (iii) the $\bar{M}$- approximate composition with respect to $\mathcal{I}$ of $\{\hat{S}_i\}_{i\in I}$ is well defined in the sense of \textit{Definition \ref{DefComposition}}. Thus, condition (iii) in \textit{Definition \ref{DefinitionBisimulation}} holds with $u^{ext}=(u_1^{ext},\ldots,u_N^{ext})$ satisfying $\mathbf d_{U^{ext}}(u^{ext},\hat{u}^{ext})=\max\limits_ {i \in I}\{\mathbf d_{U_i^{ext}}(u^{ext}_i,\hat{u}^{ext}_i)\}  = \max\limits_{i \in I}\{\mu_i\} =\mu$, and one obtains $\hat{S}_{{M}}\preccurlyeq^{\varepsilon,\mu} {S}_{\bar{M}}$. 
	
	Now from (i), using the fact that $\beta_{i}(\varepsilon,1) + \gamma^{int}_{i}(2\varepsilon+\mu_i) \leq \varepsilon,~~  \forall i \in I$, one gets from Theorem \ref{THM6} that ${S}_{\bar{M}} \preccurlyeq^{\varepsilon,0}  {S}_{\hat{M}}$. Hence, using the transitivity relation of the simulation relation in Proposition \ref{prop1}, one gets $\hat{S}_{\hat{M}}\preccurlyeq^{2\varepsilon,\mu} {S}_{{M}}$. $ \square $

	\end{pf}


 
 
	\subsection{Approximate Alternating Bisimilar Composition}
	We now present the analogous result for approximate alternating bisimulation relations.

	\begin{thm}\label{THMCompVerification}
		 Let $\{S_{i}\}_{i \in I}$ and $\{\hat{S}_{i}\}_{i \in I}$ be two collections of transition systems with $S_{i}=(X_{i}, X_{i}^{0}, U_{i}^{ext}, U_{i}^{\mathrm{int}}, \Delta_{i}$, $Y_{i}, H_{i})$ and $\hat{S}_{i}=(\hat{X}_{i}, \hat{X}_{i}^{0}, \hat{U}_{i}^{ext}, \hat{U}_{i}^{int}, \hat{\Delta}_{i}, \hat{Y}_{i}, \hat{H}_{i})$. Consider positive constants $\varepsilon_{i}, \mu_{i},$ for $ i \in I$, with $\varepsilon=\max _{i \in I} \varepsilon_{i}$, $\mu=\max_{i \in I} \mu_{i}$ and consider $M=\left(\delta_{1}, \ldots, \delta_{N}\right) $ and $\hat{M}=(\mu_1+\delta_1+\varepsilon,\ldots, \mu_N+\delta_N+\varepsilon)$ . Let the following conditions hold:
		\begin{itemize}
			\item[$(\mathrm{i})$]  For   all  $i \in I$,  $ S_{i} $ is $ \delta $-ISS and satisfies the following inequality $ \max_{i \in I} \big(\beta_{i}(\varepsilon,1) + \gamma^{int}_{i}(2\varepsilon+\mu_i)  \big)\leq \varepsilon$;
			\item[$(\mathrm{ii})$]  For   all  $i \in I$,  $ S_{i} $ is  $ (\varepsilon_{i},\mu_{i}) $- approximately alternatingly bisimilar  to  $ \hat{S}_{i} $, and we denote $ S_{i} \approx^{\varepsilon_{i},\mu_{i}} \hat{S}_{i} $;
			\item[$(\mathrm{iii})$] $\left\{S_{i}\right\}_{i \in I}$ are compatible for $M$-approximate composition with respect to $\mathcal{I}$;
			\item[$(\mathrm{iv})$] $\{\hat{S}_{i}\}_{i \in I}$ are compatible for $\hat{M}$-approximate composition with respect to $\mathcal{I}$;
		\end{itemize}
		then, ${S}_{{M}}=\left\langle {S}_{i}\right\rangle_{i \in I}^{{M}, \mathcal{I}}$ is  $(2\varepsilon,\mu)$-approximately alternatingly bisimilar to $\hat{S}_{\hat{M}}=\langle \hat{S}_{i}\rangle_{i \in I}^{\hat{M}, \mathcal{I}}$.
	\end{thm}
 \begin{pf}
           First, we have from \cite{saoud2021compositional} that $ {S}_{{M}} \preccurlyeq_{\mathcal{A}}^{\varepsilon,\mu}  \hat{S}_{\hat{{M}}}$. Hence, one gets from Proposition \ref{prop2} that ${S}_{M} \preccurlyeq_{\mathcal{A}}^{2\varepsilon,\mu}  \hat{S}_{\hat{{M}}}$. Let us now show that $\hat{S}_{\hat{M}}\preccurlyeq_{\mathcal{A}}^{2\varepsilon,\mu} {S}_{{M}}$.
		
		For $i \in I$, let $\mathcal{R}_i$ be the $(\varepsilon_i,\mu_i)$-approximate alternatingly simulation relation from $\hat{S}_i$ to $S_i$, and let us first show that the relation $\mathcal{R}$ defined by 
		$\mathcal{R}= \{ \left(x, \hat{x}\right)  \in X \times \hat{X} \: | \: (x_i,\hat{x}_i) \in \mathcal{R}_i \}$, with $x=\left(x_{1}, \ldots, x_{N}\right)$ and $\hat{x}=\left(\hat{x}_{1}, \ldots, \hat{x}_{N}\right)$, is an $(\varepsilon,\mu)$-approximate alternatingly simulation relation from $\hat{S}_{\hat{M}}$ to $S_{\bar{M}}$, with $\bar{M}=(2\mu_1+\delta_1+2\varepsilon,\ldots, 2\mu_N+\delta_N+2\varepsilon)$. 
		
		The first condition is directly satisfied (we suppose that for all $ i\in N $, $ \hat{S}_{i} $ is $(\varepsilon_i,\mu_i)$-approximately alternatingly similar to $ S_{i} $, thus, for each initial condition in $ \hat{S}_{\hat{M}} $ we can find an initial condition in $ S_{\bar{M}} $).

		Let $(x, \hat{x}) \in \mathcal{R}$ with $x=\left(x_{1}, \ldots, x_{N}\right)$ and $\hat{x}=\left(\hat{x}_{1}, \ldots, \hat{x}_{N}\right)$. Using the definition of the output map for approximate composition, relation \eqref{Eq10} and condition (ii) of \textit{Definition \ref{DefinitionAlternatingBisimulation}}, we have, 
		\begin{align*}
			&\resizebox{1\hsize}{!}{$ \mathbf{d}(H(x), \hat{H}(\hat{x}))=\mathbf{d}((H_{1}(x_{1}), \ldots, H_{N}(x_{N})),(\hat{H}_{1}(\hat{x}_{1}), \ldots, \hat{H}_{N}(\hat{x}_{N}))) $}\nonumber \\
			&=\max _{i \in I} \mathbf{d}_{Y_{i}}(H_{i}\left(x_{i}\right), \hat{H}\left(\hat{x}_{i}\right)) \leq \max _{i \in I} \varepsilon_{i}=\varepsilon
		\end{align*}
		where the last inequality follows from the fact that $(x_i,\hat{x}_i) \in \mathcal{R}_i$ for all $i \in I$.

    Consider $(x,\hat{x})\in\mathcal{R}$ with $x=(x_1,\ldots,x_N)$ and $\hat{x}=(\hat{x}_1,\ldots,\hat{x}_N)$, and any $\hat{u}^{ext} \in \hat{U}_{\hat{S}_{\hat{M}}}(\hat{x})$ with $\hat{u}^{ext}=(\hat{u}_1^{ext},\ldots,\hat{u}_N^{ext})$. Let us prove the existence of $u^{ext} \in U_{S_{\bar{M}}}(x)$ with $\mathbf{d}_{U^{ext}}(u^{ext},\hat{u}^{ext}) \leq \mu$ such that for all $\hat{x}' \in \hat{\Delta}_{\hat{M}}(\hat{x},\hat{u})$, we have the existence of ${x}' \in {\Delta}_{\bar{M}}({x},{u})$ satisfying $(x',\hat{x}')\in \mathcal{R}$.
    
	From the definition of the relation $\mathcal{R}$, we have for all $i\in I$, $(x_i,\hat{x}_i)\in \mathcal{R}_i$, then from the third condition of \textit{Definition \ref{DefinitionAlternatingBisimulation}}, we have for all $(\hat{u}_i^{ext},\hat{u}_i^{int}) \in \hat{U}_{\hat{S}_i}(\hat{x}_i)$, the existence of $(u_i^{ext},u_i^{int})\in U_{S_i}(x_i)$ with $\mathbf{d}_{U_i^{ext}}(u_i^{ext},\hat{u}_i^{ext})\leq\mu_i$ and $\mathbf{d}_{U_i^{int}}(u_i^{int},\hat{u}_i^{int})\leq\mu_i$, such that for all $x_i'\in \Delta_i(x_i,u_i^{ext},u_i^{int})$, we have the existence of $\hat{x}_i'\in \hat{\Delta}_i(\hat{x}_i,\hat{u}_i^{ext},\hat{u}_i^{int})$ satisfying $(x_i',\hat{x}_i')\in \mathcal{R}_i$.

	Let us show that the input ${u}^{int}=({u}^{int}_1,\ldots,{u}^{int}_N)$ satisfies the requirement of the $\bar{M}$-approximate composition of the components $\{{S}_i\}_{i\in I}$. The condition $\mathbf{d}_{U_i^{int}}(u_i^{int},\hat{u}_i^{int})\leq \mu_i$ implies that,
\begin{align*}
&\resizebox{1\hsize}{!}{$\mathbf{d}_{U_i^{int}}({u}_i^{int},\prod_{j\in \mathcal{N}(i)}\{{y}_j\}) \leq \mathbf{d}_{U_i^{int}}(\hat{u}_i^{int},u_i^{int})+\mathbf{d}_{U_i^{int}}(\hat{u}_i^{int},\prod_{j\in \mathcal{N}(i)}\{{y}_j\})$} \\
& \resizebox{1\hsize}{!}{$\leq \mathbf{d}_{U_i^{int}}(\hat{u}_i^{int},u_i^{int})+\mathbf{d}_{U_i^{int}}(\hat{u}_i^{int},\prod_{j\in \mathcal{N}(i)}\{\hat{y}_j\})+\mathbf{d}_{U_i^{int}}(\prod_{j\in \mathcal{N}(i)}\{y_j\},\prod_{j\in \mathcal{N}(i)}\{\hat{y}_j\})$}\\
& \resizebox{1\hsize}{!}{$\leq \mu_i + \mu_i+\delta_i+\varepsilon+ \max\limits_{j\in \mathcal{N}(i)} \varepsilon_j \leq 2\mu_i +\delta_i+\varepsilon+ \max\limits_{j \in I} \varepsilon_j =2\mu_i+\delta_i+2\varepsilon.$}
\end{align*}
	Hence, from (iii) the $\bar{M}$- approximate composition with respect to $\mathcal{I}$ of $\{\hat{S}_i\}_{i\in I}$ is well defined in the sense of \textit{Definition \ref{DefComposition}}. Thus, condition (iii) in \textit{Definition \ref{DefinitionAlternatingBisimulation}} holds with $u^{ext}=(u_1^{ext},\ldots,u_N^{ext})$ satisfying $\mathbf d_{U^{ext}}(u^{ext},\hat{u}^{ext})=\max\limits_ {i \in I}\{\mathbf d_{U_i^{ext}}(u^{ext}_i,\hat{u}^{ext}_i)\}  = \max\limits_{i \in I}\{\mu_i\} =\mu$, and one obtains $\hat{S}_{{M}}\preccurlyeq_{\mathcal{A}}^{\varepsilon,\mu} {S}_{\bar{M}}$. 
	
	Now from (i), using the fact that $\beta_{i}(\varepsilon,1) + \gamma^{int}_{i}(2\varepsilon+\mu_i) \leq \varepsilon,~~  \forall i \in I$, one gets from Theorem \ref{THM9} that ${S}_{\bar{M}} \preccurlyeq_{\mathcal{A}}^{\varepsilon,0}  {S}_{\hat{M}}$. By using the transitivity relation of the alternating simulation relation in Proposition \ref{prop1}, one gets $\hat{S}_{\hat{M}}\preccurlyeq_{\mathcal{A}}^{2\varepsilon,\mu} {S}_{{M}}$. $ \square $
 \end{pf}

	\begin{rem}
		Although there are various approaches to compute the (in)finite abstraction for interconnected systems \cite{awan2019dissipativity,rungger2016compositional,swikir2019compositional,awan2019dissipativity,saoud2021compositional,lavaei2022dissipativity}, to the best of our knowledge, only the study of \cite{tazaki2008bisimilar} developed a compositional result for the bisimulation relation. Indeed, this relation overcomes the drawback of the simulation relation concerning the existence or not of the symbolic controller based on the abstract model. The compositional result in \cite{tazaki2008bisimilar}, and  however, is limited to finite abstractions of linear subsystems, ane proposes a result of the following form: given a large-scale system consisting of interconnected components, if each subsystem is related to its abstraction by an interconnection compatible approximate bisimulation relation, then the interconnected system is related to its abstraction by an approximate bisimulation relation. In this paper, we go one step further, by showing that if each subsystem is related to its abstraction by an approximate (alternating) bisimulation relation, then the interconnected system is related to its abstraction by an approximate (alternating) bisimulation relation. Moreover, we emphasize here that the new compositional framework of this paper is suitable for different (in)finite abstractions. It yields better modularity and flexibility in the construction of symbolic models.
	\end{rem}

	\section{Case study: Traffic flow model}

	This section illustrates the engineering relevance of the proposed framework through a traffic flow example.

	\subsection{Model description and control objective}
	
	Consider the traffic flow model (\cite{saoud2021compositional}), described as:
	{\small \begin{align*}
			&\x_1(k+1)=\left(1-\frac{T v}{1.6 l}\right) \x_1(k)+5 \uu_1(k), \\
			&\x_2(k+1)=\frac{T v}{l} \x_1(k)+\left(1-\frac{T v}{l}-q\right) \x_2(k)+\frac{T v}{l} \x_4(k), \\
			&\x_3(k+1)=\frac{T v}{l} \x_2(k)+\left(1-\frac{T v}{l}-q\right) \x_3(k)+8 \uu_3(k), \\
			&\x_4(k+1)=\frac{T v}{l} \x_3(k)+\left(1-\frac{T v}{l}-q\right) \x_4(k)+8 \uu_4(k), \\
			&\x_5(k+1)=\frac{T v}{l} \x_4(k)+\left(1-\frac{T v}{l}-q\right) \x_5(k)+8 \uu_5(k),
	\end{align*}}
	where the state $\x_i(k), i \in I=\{1,2,3,4,5\}$, represents the traffic density in the $i^{t h}$ road section, expressed in vehicles per section, $l=0.25 \mathrm{~km}$ is the length of the road, $v=70 \mathrm{~km} / \mathrm{hr}$ is the flow speed, $T=\frac{10}{3600}$ hours is the discrete-time interval, and $q=0.25$ is the ratio representing the percentage of vehicles leaving the section of road. For each $ i^{th} $ difference equation, the states $ x_{j} $ with $ j \neq i,  \; i=\{1,2,3,4,5\} $ represent the internal inputs. The external inputs $\uu_1(k), \uu_3(k), \uu_4(k), \uu_5(k) \in U=\{0,1\}$, where $ 0 $ represents the red signal, and $ 1 $ represents the green signal in the traffic model.  We consider the compact state-space $X=[0,40]^5$. The control objective is to synthesize controller to stay inside a safe region $\mathfrak{S}=[2,25]\times[5,25]^4$. 
	
	\begin{figure}[ht]
		\centering
		\includegraphics[width=0.36\textwidth]{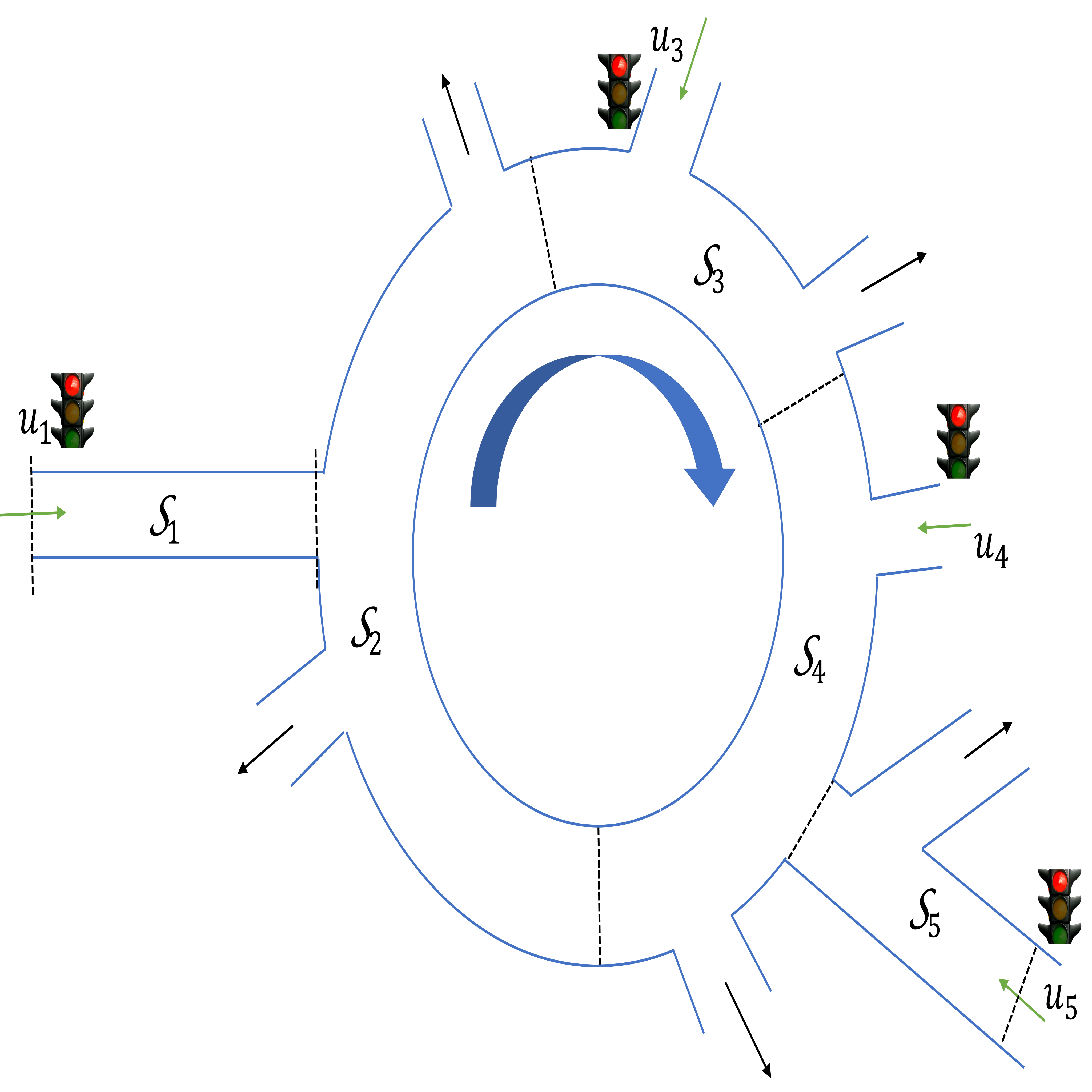}
		\caption{Traffic flow network where the clockwise flow of traffic is allowed and $\mathcal{S}_i$ represents road sections.}
	\end{figure}
	
	The proposed model can be seen as an exact composition of $5$ subsystems $S=\left\langle S_{i}\right\rangle_{i \in I}^{\mathbf{0}_{5}, \mathcal{I}}$, with, $$ \resizebox{1\hsize}{!}{$\mathcal{I} =\{\hspace{-0.1em}(1,\hspace{-0.1em}1),\hspace{-0.1em}(1,\hspace{-0.1em}2),\hspace{-0.1em}(2,\hspace{-0.1em}2),\hspace{-0.1em}(2,\hspace{-0.1em}4),\hspace{-0.1em}(3,\hspace{-0.1em}3),\hspace{-0.1em}(2,\hspace{-0.1em}3),\hspace{-0.1em}(4,\hspace{-0.1em}4),\hspace{-0.1em}(3,\hspace{-0.1em}4),\hspace{-0.1em}(5,\hspace{-0.1em}5),\hspace{-0.1em}(4,\hspace{-0.1em}5)\hspace{-0.1em}\}$}.$$
	
	\subsection{Abstraction and controller synthesis}
	
	First one can check that each subsystem $S_i$, $i \in I$ is $\delta$-ISS with $\beta_1(r,k)=(0.513)^k r$, $\beta_2(r,k)=\beta_3(r,k)=\beta_4(r,k)=\beta_5(r,k)=(0.0287)^k s$, $\gamma^{\textit{int}}_1(r)=0.01r$ and $\gamma^{\textit{int}}_2(r)=\gamma^{\textit{int}}_3(r)=\gamma^{\textit{int}}_4(r)=\gamma^{\textit{int}}_5(r)=0.195 r$ .
	We compute local abstraction $\hat{S}_i$ for each subsystem $S_i$, $i \in I$,  using the symbolic approach presented in~\cite{girard2009approximately}. Each abstraction $\hat{S}_i$ is related to the original system $S_i$, $i\in I$, by an $(\varepsilon_i,\mu_i)$-approximate bisimulation relation, with $\varepsilon_i=1$ and $\mu_i=1$. We then compose the local abstractions in order to compute the global abstraction using an $\hat{M}$-approximate composition, with $\hat{M}=(1,1,1,1,1)$.  One can also check that for the chosen values of $\varepsilon_i$ and $\mu_i$, $i \in I$, condition (i) of Theorem~\ref{THMCompVerification} is satisfied. Hence, in view of Theorem~\ref{THMCompVerification}, we have that $\hat{S}\approx_{\mathcal{A}}^{(\varepsilon,\mu^{\prime})}S$ \footnote{Given the safety specification for the original system $\mathfrak{S}$ and since the original system is related to the compositional abstraction by an $\varepsilon-$approximate bisimulation relation, the abstract specification is a deflated version of the original one.}, where $S=\langle S^i \rangle_{i\in I}^{\mathbf{0}_4,\mathcal{I}}$ and $\hat{S}=\langle \hat{S}_i \rangle_{i\in I}^{\hat{M},\mathcal{I}}$.
	
	The computation time of the abstractions of the four components $\{1,2,3,4,5\}$ are given by $0.22$ seconds, $0.25$ seconds, $0.16$ seconds, $0.14$ seconds and $0.15$ seconds, respectively, and the composition of the global abstraction from local ones using an approximate composition takes less than $138$ seconds. This resulted in $139$ seconds to compute an abstraction compositionally. Constructing an abstraction for the full model monolithically, using the same discretization parameters, took $241$ seconds. Hence, the proposed compositional approach is two times faster in this scenario.

\begin{figure}[t!]
		\centering
		\includegraphics[scale=0.5]{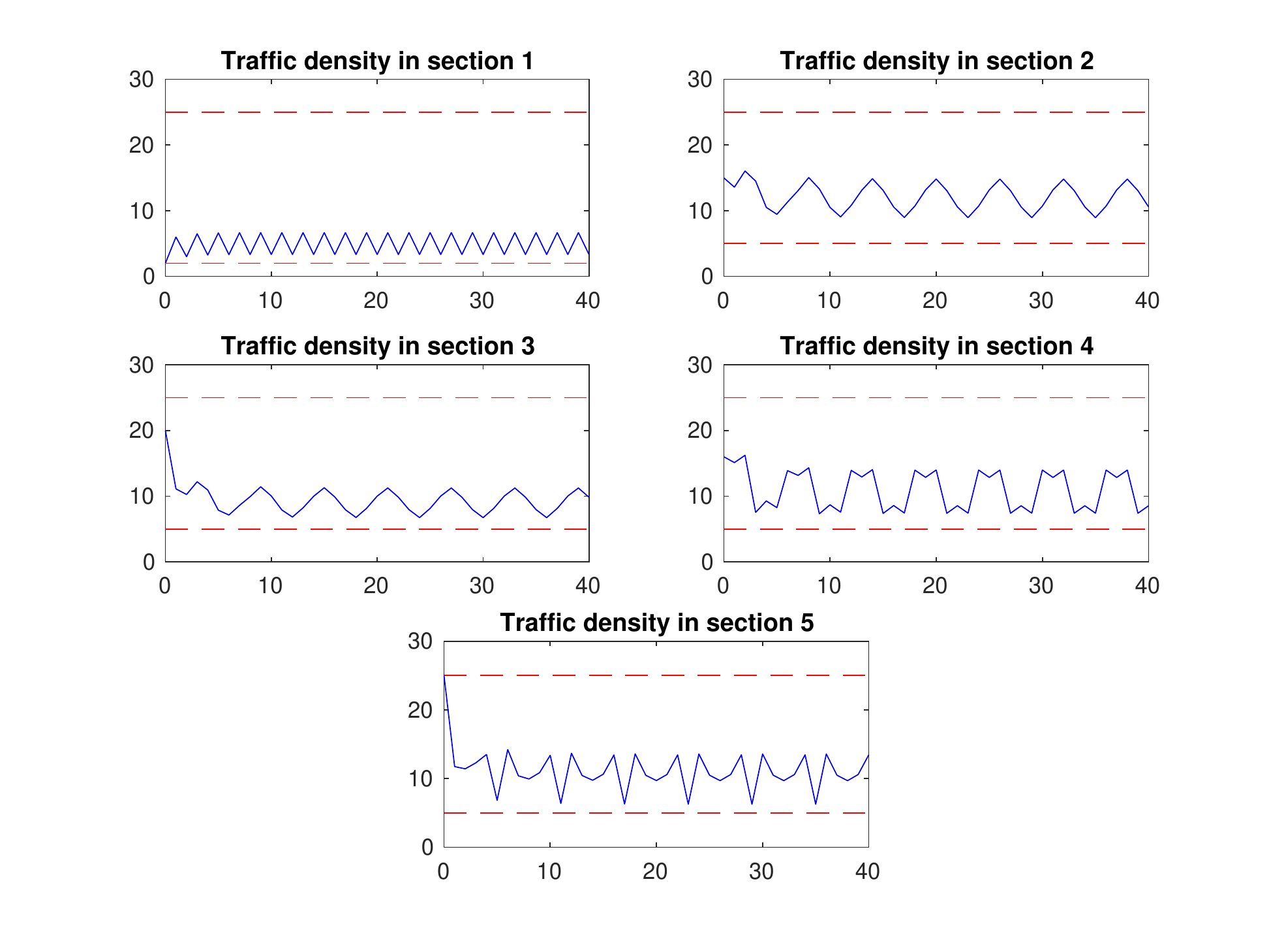}\vspace{-0.6cm} \caption{The evolution of traffic densities in each section of the road.}
		\label{fig:resp_traffic}
		\vspace{-0.1cm}
	\end{figure}

	Figure \ref{fig:resp_traffic} shows the evolution of traffic densities in each section of the road starting from the initial condition $x=[2,15,20,16,25]$ using a safety controller synthesized for the constructed compositional abstraction. The dashed red lines represent the boundary of the safe set for each section. One can readily see that all the trajectories evolve within the safe region.   

	\section{Conclusion}
	This paper studied the problem of abstraction of interconnected transition systems. A compositional framework for constructing abstractions is proposed based on the notion of approximate composition and the $ \delta-ISS $ properties. In particular, given a large-scale system consisting of interconnected components, we provided conditions under which the concept of approximate (alternating) simulation relation is preserved when going from the subsystems to the large-scale interconnected system. A numerical result is proposed showing the merits of the theoretical results. 

	\bibliography{ifacconf}             
	
	
	
	
	
	
	
	
	\appendix
	
	\section{Proofs}
	\subsection{Proof of Proposition 1}  \label{A1}  
	Although the structure of this proof follows similar steps to proof \textit{Proposition 2.9} presented in \cite{juluis2006}, it differs from two pints. Firstly, $ S_{1}, \; S_{2} $ and $ S_{3} $ are interconnected transition systems as defined in \textit{Definition \ref{DefTransitionSys}} and second, the approximate simulation is as in \textit{Definition \ref{DefinitionBisimulation}}. We note also that the second item regarding alternating simulation relation follow similar steps. \\
	Let the relation $\mathcal{R}_{12}$ defined by $S_{1} \preccurlyeq_{\varepsilon, \mu} S_{2}$ and the relation $\mathcal{R}_{23}$  defined by $S_{2} \preccurlyeq_{\varepsilon^{\prime}, \mu^{\prime}} S_{3}$. The first relation defines an $(\varepsilon, \mu)$ -- approximate simulation of $T_1$ by $T_2$, and the second defines an $\left(\varepsilon^{\prime}, \mu^{\prime}\right)$ -- approximate simulation of $T_{2}$ by $T_{3}$. The aim is to prove that,
	\begin{align}\label{ADIXEq1}
		&\mathcal{R}_{13} :=\mathcal{R}_{12} \circ \mathcal{R}_{23},\nonumber\\ 
		&=\left\{\left(q_1, q_3\right) \mid \exists q_2,\left(q_1, q_2\right) \in \mathcal{R}_{12},\left(q_2, q_3\right) \in \mathcal{R}_{23}\right\}
	\end{align}
	is a $\left(\varepsilon+\varepsilon^{\prime}, \mu+\mu^{\prime}\right)$ -- approximate simulation of $T_2$ by $T_3$. Choose any $\left(q_1, q_3\right) \in \mathcal{R}_{23}$. First, we show that
	\begin{align}\label{ADIXEq2}
		\mathbf{d}\left( q_1, q_3\right) \leq \varepsilon+\varepsilon^{\prime}.
	\end{align}
	By definition of $\mathcal{R}_{13}$, there exists a $q_2 \in Q_2$ such that $\left(q_1, q_2\right) \in \mathcal{R}_{12}$ and $\left(q_2, q_3\right) \in \mathcal{R}_{23}$. From there, we can deduce that $ \exists q_{2}  $ satisfying \eqref{ADIXEq1} such that,
	\begin{align}
		\left\lbrace \begin{array}{c}
			\mathbf{d}\left( q_1, q_2\right) \leq \varepsilon \\
			\mathbf{d}\left( q_2, q_3\right) \leq \varepsilon^{\prime}
		\end{array}	\right.	& \Longrightarrow \mathbf{d}\left( q_1, q_2\right)+ \mathbf{d}\left( q_2, q_3\right) \leq \varepsilon + \varepsilon^{\prime}, \nonumber \\
		& \Longrightarrow \mathbf{d}\left( q_1, q_3\right) \leq \varepsilon + \varepsilon^{\prime},
	\end{align}
	which satisfies equation \eqref{ADIXEq2}. Now, the aim to show that if $q_1 \stackrel{u^{ext}}{\rightarrow} q_1^{\prime}$ for some $u^{ext} \in U^{ext}$ and $q_1^{\prime} \in Q_1$, then there exist $(u^{ext})^{\prime} \in U^{ext}$ and $q_3^{\prime} \in Q_3$ such that
	\begin{align}\label{ADIXEq3}
		\left(q_1^{\prime}, q_3^{\prime}\right) \in \mathcal{R}_{23}, q_3 \stackrel{(u^{ext})^{\prime}}{\rightarrow} q_3^{\prime},  \mathbf{d}_{U^{ext}}\left(u^{ext}, (u^{ext})^{\prime}\right) \leq \mu+\mu^{\prime} .
	\end{align}
	By the existence of a $q_2 \in Q_2$ as above, we deduce the existence of a $q_2^{\prime} \in Q_2$ and $(u^{ext})^{\prime \prime} \in U^{ext}$ such that,
	\begin{align}\label{ADIXEq6}
		(q_1^{\prime}, q_2^{\prime})\hspace{-0.2em} \in\hspace{-0.2em} \mathcal{R}_{12}, q_2\hspace{-0.2em} \stackrel{(u^{ext})^{\prime \prime}}{\rightarrow} \hspace{-0.2em}q_2^{\prime}, \mathbf{d}_{U^{ext}}(u^{ext}, (u^{ext})^{\prime \prime}) \hspace{-0.2em}\leq\hspace{-0.2em} \mu .
	\end{align}
	This in turn implies the existence of a $q_3^{\prime} \in Q_3$ and $({u}^{ext})^{\prime} \in U^{ext}$ such that
	\begin{align}\label{ADIXEq7}
		(q_2^{\prime}, q_3^{\prime})\hspace{-0.2em} \in\hspace{-0.2em} \mathcal{R}_{23}, q_3 \hspace{-0.2em}\stackrel{(u^{ext})^{\prime}}{\rightarrow} \hspace{-0.2em}q_3^{\prime}, \mathbf{d}_{U^{ext}}(\hspace{-0.1em}(\hspace{-0.1em}u^{ext})^{\prime}\hspace{-0.2em}, \hspace{-0.2em}(\hspace{-0.1em}u^{ext})^{\prime \prime}\hspace{-0.1em}) \hspace{-0.2em}\leq\hspace{-0.2em} \mu^{\prime} .
	\end{align}
	Notice that one can obtain \eqref{ADIXEq3} by adding  \eqref{ADIXEq6} and \eqref{ADIXEq7}. 
	
	\subsection{Proof of Proposition 2}     \label{A2}         
	
	Consider two pseudometric transition systems $S_{1}$ and $S_{2}$ satisfying \textit{Definition \ref{DefTransitionSys}}. The aim is to prove that if the relation $ \mathcal{R}_{1} $ defined by, $S_{1} \preccurlyeq_{\varepsilon, \mu} S_{2}$ holds, then, for $ \mu^{\prime} \geq \mu  $ and $\varepsilon^{\prime} \geq \varepsilon$, the relation $\mathcal{R}_{2} $ defined by $S_{1} \preccurlyeq_{\varepsilon^{\prime}, \mu^{\prime}} S_{2}$ holds.	
	
	Suppose that we have the relation $S_{1} \preccurlyeq_{\varepsilon, \mu} S_{2}$ with $ \mu \geq 0$ and $ \varepsilon \geq 0$. Due to the relation $S_{1} \preccurlyeq_{\varepsilon, \mu} S_{2}$, the three conditions in \textit{Definition \ref{DefinitionBisimulation}} are satisfied, whereas the second and third conditions are satisfied with the constants $ \mu $ and $ \varepsilon $, respectively.  Now, we want to prove the relation $ S_{1} \preccurlyeq_{\varepsilon^{\prime}, \mu^{\prime}} S_{2} $. The first condition in \textit{Definition \ref{DefinitionBisimulation}}, is directly satisfied. Since $ \mu^{\prime}\geq \mu $ and $ \varepsilon^{\prime} \geq \varepsilon $, the second and the third conditions follows because of the pseudometric properties.	 
    The second item  the alternating simulation relation follows similar steps.

 \subsection{Proof of Theorem \ref{THM1}}
 \label{A3}
    Let $ \x^{a}$ and $\x^{b} $ be two trajectories of the system \eqref{LinearSys}. At time $ k+1 $, the difference between $ \x^{a}$ and $\x^{b} $ is defined by,
	$ \x^{a}(k+1)-\x^{b}(k+1)=A (\x^{a}(k)-\x^{b}(k)) 
	+B(\uu^{ext,a}(k)-\uu^{ext,b}(k))  
	+D(\uu^{int,a}(k)-\uu^{b,int}(k)).$
	Now, we can rewrite it as,
	$\x^{a}(k+1)-\x^{b}(k+1)=A^{k+1} (\x^{a}(0)-\x^{b}(0))
	+\sum_{j=0}^k A^{k-j}B (\uu^{ext,a}(j)-\uu^{ext,b}(j))
	+\sum_{j=0}^k A^{k-j}D(\uu^{int,a}(j)-\uu^{int,b}(j)).$
	It can be concluded that, if all the eigenvalues of the matrix $ A $ are inside the unite disk, then, the $ \beta(r, k) $ in \textit{Theorem \ref{THM1}} is deceasing with respect to its second argument. This implies that the $ \delta-ISS $ property is satisfied and the functions $ \beta(r, k) ,\gamma^{ext}(r)$ and $ \gamma^{int}(r) $ are defined as in Theorem \ref{THM1}. 

	\subsection{Proof of Theorem \ref{THM2}}
	\label{A4}
	Let $ \x^{a}$ and $\x^{b} $ be two trajectories of the system \eqref{LinearSys}. At time $ k+1 $, the difference between $ \x^{a}$ and $\x^{b} $ is defined by,
	$ \x^{a}(k+1)-\x^{b}(k+1)=f(\x^{a}(k),\textbf{u}^{ext,a}(k),\textbf{u}^{int,a(k})-f(\x^{b}(k),\textbf{u}^{ext,b}(k),\textbf{u}^{int,b}(k))$. Thanks to the Lipschitz condition \eqref{Eq28}, we have, $ \x^{a}(k+1)-\x^{b}(k+1)\leq L^{x}\left\|\x^{a}(k)-\x^{b}(k)\right\|+ L^{u^{ext}}\left\|\uu^{ext,a}(k)-\uu^{ext,b}(k)\right\|+ L^{u^{int}}\left\|\uu^{int,a}(k)-\uu^{int,b}(k)\right\|$. 
	Now, we can rewrite it as, $\x^{a}(k+1)-\x^{b}(k+1)\leq (L^{x})^{k+1} (\x^{a}(0)-\x^{b}(0)) 
	+\sum_{j=0}^k (L^{x})^{k-j}L^{u^{ext}} (\textbf{u}^{ext,a}(j)-\textbf{u}^{ext,b}(j)) 	 
$ \\ $ +\sum_{j=0}^k  (L^{x})^{k-j} (L^{u^{int}})(\textbf{u}^{int,a}(j)-\textbf{u}^{int,b}(j))$.
	It can be concluded that, if $ L^{x} <1 $, then, the $ \beta(r, k) $ in \textit{Theorem \ref{THM2}} is deceasing with respect to its second argument. This implies that the $ \delta-ISS $ property is satisfied and the functions $ \beta(r, k) ,\gamma^{ext}(r)$ and $ \gamma^{int}(r) $ are defined as in \textit{Theorem 
 \ref{THM2}}. 	
	
\end{document}